\documentclass[10pt,aps,prd,amsmath,amssymb,superscriptaddress,onecolumn,nofootinbib,showpacs,preprintnumbers,amsfonts, notitlepage]{revtex4-2}

\usepackage{hyperref}
\usepackage{amsmath,amssymb}
\usepackage{bm}
\usepackage{multirow}
\usepackage{xspace}
\usepackage{colortbl}
\usepackage{tikz}
\DeclareMathOperator{\im}{Im}

\newcommand{\ie}{{\em i.e.}\xspace}
\usepackage{pifont}
\newcommand{\cmark}{\ding{51}}%
\newcommand{\xmark}{\ding{55}}%

\usepackage[compat=1.1.0]{tikz-feynman}

\providecommand{\U}[1]{\protect\rule{.1in}{.1in}}
\newcommand{\gev}{\ensuremath{{\mathrm{\,Ge\kern -0.1em V}}}\xspace}
\newcommand{\gevsq}{\ensuremath{{\mathrm{\,Ge\kern -0.1em V}}^2}\xspace}
\newcommand{\mev}{\ensuremath{{\mathrm{\,Me\kern -0.1em V}}}\xspace}

\begin{document}

\title{Glueball-glueball scattering and the glueballonium}

\newcommand{\kielce}{Institute of Physics, Jan Kochanowski University, ul. Uniwersytecka 7, 25-406, Kielce, Poland}
\newcommand{\frankfurt}{Institute for Theoretical Physics, J. W. Goethe University, Max-von-Laue-Str. 1, 60438 Frankfurt, Germany}
\newcommand{\catania}{INFN Sezione di Catania, Via Santa Sofia 64, I-95123 Catania, Italy}
\newcommand{\mift}{Dipartimento di Scienze Matematiche e Informatiche, Scienze Fisiche e Scienze della Terra,
Universit\`a degli Studi di Messina, I-98166 Messina, Italy}

\author{Francesco~Giacosa}
\affiliation{\kielce}
\affiliation{\frankfurt}

\author{Alessandro~Pilloni}
\affiliation{\mift}
\affiliation{\catania}

\author{Enrico~Trotti}
\affiliation{\kielce}

\begin{abstract}
The scalar glueball $G$ is the lightest particle of the
Yang-Mills sector of QCD, with a lattice predicted mass of about
$m_{G}\simeq1.7\gev$. It is natural to investigate glueball-glueball
scattering and the possible emergence of a bound state,
that we call glueballonium. We perform this study in the context of a
 widely used dilaton potential, that depends on a single dimensionful parameter $\Lambda_G$. We consider a unitarization prescription that allows us to predict the lowest partial waves in the elastic window. These quantities can be in principle calculated on the
lattice, thus offering possibility for testing the validity of the dilaton
potential and an independent determination of its parameter. Moreover, we
also show that a stable glueballonium exists if $\Lambda_{G}$ is small enough. In particular, for 
$\Lambda_{G}$ compatible with the expectations from the gluon condensate, the glueballonium has a mass of about $3.4\gev$.
\end{abstract}

\maketitle

\section{Introduction}

Glueballs, bound states of gluons, are firm predictions of the Yang-Mills (YM) sector of
Quantum Chromodynamics (QCD). They were originally proposed within bag models~\cite{Chodos:1974je,Jaffe:1975fd,Jaffe:1985qp}, and later confirmed by Lattice QCD calculations~\cite{Sharpe:1998hh,Morningstar:1999rf,Chen:2005mg,Gregory:2012hu,Caselle:2001im,Athenodorou:2020ani}.
Other nonperturbative approaches lead to similar conclusions~\cite{Ochs:2013gi,Mathieu:2008me,Crede:2008vw,Dosch:2002hc,Forkel:2003mk,Szczepaniak:2003mr,Isgur:1985vy,Brower:2000rp,Sanchis-Alepuz:2015hma,Gounaris:1985uy,Vento:2017ice,Huber:2020ngt}. 
All these works agree that the lightest degree of freedom of pure YM is a scalar glueball $G$
with a mass of about $m_{G}\simeq1.7\gev$.
The scattering of two glueballs is therefore a well defined
process in pure YM, that can be investigated both by models and
on the lattice by using the
L\"{u}scher method \cite{Luscher:1990ux,Andersen:2018mau} (for a first exploratory lattice study see~\cite{Yamanaka:2019gak,Yamanaka:2019yek,Yamanaka:2019aeq,Yamanaka:2021xqh}). 
Besides masses, the behavior of the phase shifts delivers valuable information to understand the nonperturbative nature of glueballs and their interactions. 

In this work, we study the scattering of two scalar glueballs in the framework of the well-known dilaton
potential, which (in its simplest form) involves  a single dilaton/glueball scalar field $G$~\cite{Migdal:1982jp,Salomone:1980sp,Gomm:1984zq,Gomm:1985ut}.
This potential represents an effective description of a fundamental property of YM called the trace anomaly, according to which the a low-energy scale $\Lambda_\text{YM}\simeq 250\mev$
is dynamically generated due to gluonic quantum fluctuations and to the gluonic condensate~\cite{Thomas:2001kw,Shifman:1988zk}. In QCD, this feature is ultimately connected to the masses of hadrons, since each mass is proportional to $\Lambda_\text{YM}$ if light quark masses are neglected. 
In the context of the dilaton potential, there is an analogous dimensionful constant $\Lambda_{G}$  proportional to $N_{c}\Lambda_\text{YM},$ where $N_{c}$
is the number of colors (3 in Nature). 

In this paper, we address the following  questions: $(i)$ What are the scattering amplitudes within the dilaton potential? $(ii)$ Is the attraction between two glueballs strong enough to generate a bound state?  Quite interestingly, for values of the parameter in reasonably agreement with lattice expectations, such a glueball-glueball state, that we call glueballonium, might exist. If so, it is stable in YM, and could appear as on the lattice as an excited scalar glueball with a mass of about $3\gev$.

Moreover, the interaction between two glueballs is also of primary importance for full QCD. In the past, various states were proposed as  predominantly gluonic candidates, most notably  the $f_{0}(1500)$ and $f_{0}(1710)$~\cite{Janowski:2014ppa,Amsler:1995td,Close:2001ga,Brunner:2015oqa,Brunner:2015yha,Lee:1999kv,Gui:2012gx,Giacosa:2005qr,Cheng:2006hu,Rodas:2021tyb},\footnote{Recently, Ref.~\cite{Klempt:2021nuf} proposed the possibility that hints of the glueball are equally spread in the scalar-isoscalar sector.} yet all the discussed scenarios are meaningful under the assumption that the scalar glueball is narrow enough. 
Quite remarkably, it turns out the very decay of the scalar glueball into pions (as well as into other mesons) depends in ultimate analysis on the glueball-glueball interaction strength parametrized by the parameter $\Lambda_{G}$ mentioned above. In particular, as already pointed out in Ref.~\cite{Ellis:1984jv}, the scalar glueball might be broader than $1\gev$, thus making its experimental discovery very hard, if not impossible. 
In this respect, the understanding of  glueball-glueball scattering in pure YM is relevant to answer the question about the width of the light scalar glueball, which is crucial for its possible discovery. 
The glueballonium itself, if exists in pure YM, could also appear as a resonance at about 3\gev, if the constituent glueballs are not too broad.

Furthermore, understanding the dilaton potential is important on its own, since it is often part of QCD models that contain
quark-antiquark mesons~\cite{Heide:1993yz,Carter:1995zi,Fariborz:2018xxq,Fariborz:2018unf,Parganlija:2012fy,Parganlija:2010fz,Gutsche:2012ez} and affects ---directly and indirectly--- the decay of ordinary mesons~\cite{Janowski:2014ppa,Janowski:2011gt}, as well as the behavior of QCD
 at nonzero temperature and densities~\cite{Drago:2001gd,Papazoglou:1997uw,Schaefer:2001cn,Agasian:2008zza,Park:2008zg,Paeng:2011hy,Kovacs:2016juc}. The
dilaton potential might also be relevant beyond QCD~\cite{Goldberger:2007zk,Tseytlin:1991ss,Gasperini:2007ar}.

The article is organized as follows: in Sec.~\ref{sec:theory} we recall the main features of the
dilaton potential, in Sec.~\ref{sec:tree} we present the scattering amplitudes at tree-level, which we unitarize in Sec.~\ref{sec:unitarization} introducing a suitable
scheme and studying the
emergence of the glueballonium. In Sec.~\ref{sec:heavy} we show that heavy glueballs (both
scalar and non-scalar) do not affect the results presented in the previous sections. Finally, in Sec.~\ref{sec:conclusions} we present our conclusions.

\section{From YM to the effective theory of glueballs}
\label{sec:theory}

\noindent
We briefly recall how the dilaton/glueball field $G$ emerges
as a low-energy theory of the YM Lagrangian. The latter depends on a single
dimensionless coupling $g_{0}$:
\begin{align}%
\mathcal{L}%
_{\text{YM}} & =-\frac{1}{4}G_{\mu\nu}^{a}G^{a,\mu\nu} &\text{with }G_{\mu\nu}%
^{a}&=\partial_{\mu}A_{\nu}^{a}-\partial_{\nu}A_{\mu}^{a}+g_{0}f^{abc}A_{\mu
}^{b}A_{\nu}^{c}\text{ ,}\label{ymlag}%
\end{align}
where $G^{a,\mu\nu}$ is the gluon field-strength tensor, $A_{\mu}^{a}$ is the
gluon field with $a=1,\ldots,N_{c}^{2}-1$, $f^{abc}$ are the $SU(N_c)$ structure constants.

Although the YM Lagrangian is classically invariant under dilatation transformations,
$x^{\mu}\rightarrow\lambda^{-1}x^{\mu}$ and $A_{\mu}^{a}(x)\rightarrow\lambda
A_{\mu}^{a}(\lambda x)$, this symmetry is broken by quantum
fluctuations. This is the famous trace anomaly, a basic property of non-abelian
gauge theories. Upon renormalization, the coupling constant $g_{0}$ becomes a function $g(\mu)$
of the energy scale $\mu$. As a consequence, the divergence of the dilatation
current no longer vanishes~\cite{Thomas:2001kw,Shifman:1988zk,Donoghue:1992dd}:
\begin{equation}
\partial_{\mu}J_{\text{dil}}^{\mu}=T_{\mu}^{\mu}=\frac{\beta
(g)}{2g}\,G_{\mu\nu}^{a}G^{a,\mu\nu}\neq0\text{ ,}\label{ta}%
\end{equation}
where $\beta(g)=\partial g/\partial\ln\mu$ and $T^{\mu\nu}$ the symmetric energy-momentum tensor of the YM
Lagrangian. At one-loop $\beta(g)=-bg^{3}$ with $b=11N_{c}/(48\pi^{2})$,
then:
\begin{equation}
g^{2}(\mu)=\frac{1}{2b\ln(\mu/\Lambda_{\text{YM}})}\,,
\end{equation}
where $\Lambda_\text{YM}$ is the YM scale, which realizes the so-called
dimensional transmutation. Notice that in massless QCD any dimensionful quantity, such as the hadron masses, must be
proportional to $\Lambda_\text{YM}$ even though we are presently unable to determine the proportionality constant analytically.

Moreover, the nonvanishing expectation value of the trace anomaly reads:%
\begin{equation}
\left\langle T_{\mu}^{\mu}\right\rangle =-\frac{11N_{c}}%
{24}\left\langle \frac{\alpha_{s}}{\pi}\,G_{\mu\nu}^{a}G^{a,\mu\nu
}\right\rangle =-\frac{11N_{c}}{24}C^{4}\,,\label{gc}%
\end{equation}
 with $C^4 = \left\langle \frac{\alpha_{s}}{\pi}\,G_{\mu\nu}^{a}G^{a,\mu\nu
}\right\rangle $ being the gluon condensate. For the relevant $N_c=3$ case,
\begin{equation}
(C^{4})_{N_c=3} \approx (0.3\text{--}0.6\gev)^{4}\,,%
\end{equation}
where the numerical values are obtained in QCD sum rules (lower range of the
interval)~\cite{Narison:2014wqa,Ioffe:2005ym} and lattice YM simulations (higher range of
the interval)~\cite{DiGiacomo:2000irz}.
Note, taking into account that $\alpha_{s}$ scales as $N_{c}^{-1}$ for fixed `t~Hooft coupling, and that the trace over the adjoint index scales as $N_c^2$ in the large-$N_c$ limit, it follows that the quantity $C \equiv C(N_c)$ scales as
$N_{c}^{1/4}$.

Because of confinement, gluons are not the asymptotic states of the
theory. As said, it is universally accepted that the lightest excitation is a scalar glueball.
It is then
natural to consider a single scalar field $G$ that describes both the scalar
glueball and the trace anomaly at the hadron level.
The corresponding
effective dilaton Lagrangian reads~\cite{Migdal:1982jp,Salomone:1980sp,Gomm:1984zq,Gomm:1985ut}:
\begin{equation}%
\mathcal{L}%
_\text{dil}=\frac{1}{2}(\partial_{\mu}G)^{2}-V(G),
\end{equation}
with%
\begin{equation}
V(G)=\frac{1}{4}\frac{m_{G}^{2}}{\Lambda_{G}^{2}}\left(  G^{4}\ln\left\vert
\frac{G}{\Lambda_{G}}\right\vert -\frac{G^{4}}{4}\right)  \text{.}%
\label{potential}%
\end{equation}
The dilaton potential contains two parameters: the dimensionful one $\Lambda_G$ and the dimensionless ratio $m_G/\Lambda_G$. Once $m_G$ is fixed, it depends only on $\Lambda_G$.
The minimum of the dilaton potential is realized for $G=\Lambda
_{G}$. After the shift $G\rightarrow \Lambda_G+G$, it is easy to identify 
$m_{G}$ as the glueball mass. The numerical value of $m_{G}\approx1.7\gev$ can
be found in Lattice QCD~\cite{Morningstar:1999rf,Chen:2005mg}, as well as in diverse phenomenological studies~\cite{Janowski:2014ppa,Amsler:1995td,Close:2001ga,Brunner:2015oqa,Brunner:2015yha,Lee:1999kv,Gui:2012gx,Giacosa:2005zt,Cheng:2006hu}. 
The invariance under dilatation transformation is explicitly broken by the
logarithmic term of the potential. The divergence of the corresponding current
reads:
\begin{equation}
\partial_{\mu}J_\text{dil}^{\mu}=T_{\text{imp},\,\mu}^{\;\mu}=4V- G\partial_{G}V=-\frac
{1}{4}\,\frac{m_{G}^{2}}{\Lambda_{G}^{2}}\,G^{4}\rightarrow-\frac{1}{4}%
m_{G}^{2}\Lambda_{G}^{2} \text{ ,}\label{dil1}%
\end{equation} 
where $T_\text{imp}$ is the improved energy-momentum tensor~\cite{Callan:1970ze}.

Interestingly, by reversing the line of arguments, the
equation in Eq.~\eqref{dil1} can be used to justify the form of the dilaton
potential. 
Namely, by imposing that a certain scalar theory with a generic
potential $V(G)$ fulfills the equation $\partial_{\mu}J_{dil}^{\mu}%
=G\partial_{G}V-4V\propto G^{4}$ (thus reproducing the scale anomaly of Eq.~\eqref{ta}) one gets:
\begin{equation}
\partial_{G}V-\frac{4}{G}V=\alpha G^{3}\text{ ,}%
\end{equation}
where $\alpha$ is a dimensionless constant. The general solution is 
\begin{equation}
V(G)=\alpha G^{4}\left[  \ln\frac{G}{\Lambda_{G}}-\frac{1}{4}\right]~,
\end{equation}
with $\Lambda_G$ a (dimensionful) integration constant. The factor $1/4$ is introduced, so that the minimum is for $G=\Lambda_G$.
Finally, when introducing the mass $m_{G}^{2}$ as the curvature at the
minimum, one gets $\alpha=\frac{1}{4}\frac{m_{G}^{2}}{\Lambda_{G}^{2}}$, that matches Eq.~\eqref{potential}.

The large-$N_{c}$ behavior of the parameters is
obtained by imposing that the quartic coupling $\alpha$ is of order $N_{c}^{-2}$ and that the glueball mass is $N_{c}^0$~\cite{Witten:1979kh,Lebed:1998st}: one thus gets $\Lambda_{G}\propto N_{c}$.


The requirement that the dilaton field saturates the trace of the dilatation
current means to equate Eq.~\eqref{gc} with Eq.~\eqref{dil1} (valid for any $N_c$):
\begin{equation}
\Lambda_{G}\overset{!}{=}\sqrt{\frac{11N_c}{6}}\frac{C^2}{m_G} \text{ .}
\end{equation}

\noindent Upon using $m_{G}\approx1.7\gev$ as well as, for $N_c=3$, the value $C_{N_c=3}=0.55 \gev$,
one obtains $(\Lambda_{G})_{N_c =3}\approx0.5 \gev$. 
At the hadron level, $\Lambda_{G}$ should be the only dimensionful parameter if quark masses are neglected, as realized in
\cite{Janowski:2014ppa,Parganlija:2012fy,Parganlija:2010fz}. 
Notice that the scalings $\Lambda_{G}\propto N_{c}$ and $m_{G}\propto N_{c}^{0}$ are consistent
with the scaling $C \propto N_{c}^{1/4}$ calculated above.

In full QCD, the scalar glueball is no longer stable since it decays into light mesons such as pions and kaons. Moreover, mixing with scalar-isoscalar mesons takes place.  
These features have accompanied the study of
glueballs in full QCD as well as their experimental determination~\cite{Mathieu:2008me,Ochs:2013gi,Llanes-Estrada:2021evz,Sarantsev:2021ein,Klempt:2021nuf,Rodas:2021tyb}. 
As discussed in~\cite{Ellis:1984jv}, the value of $\Lambda_{G}$ entering into the dilaton potential affects directly the
decay width of the scalar glueball into mesons. 
In particular, a value
of $\Lambda_{G}\approx0.4\gev$ would imply that the glueball has a width of $\sim 1\gev$, too broad
to be measured~\cite{Gounaris:1985uy}.

In Appendix~\ref{app:decays} we present a toy model that elucidates how the constant  $\Lambda_{G}$ enters into the decays of the glueball. In particular, it is visible that the decay of the field $G$ into two pions scales as $\Lambda_G^{-2}$, thus the smaller $\Lambda_{G}$, the larger the decay width. 
According to Ref.~\cite{Janowski:2014ppa}, in which a realistic version of the toy
model phenomenology of Eq.~\ref{toy} is developed and a proper treatment of
mixing is worked out, a narrow glueball is realized if $\Lambda_{G}\gtrsim1 \gev$. Yet, as discussed in Ref.~\cite{Lakaschus:2018rki}, the effect of mixing with the
lightest $f_{0}(500)/\sigma$ can be relevant, reducing $\Lambda_{G}$ to about $0.5$--$1\gev$ to have a satisfactory phenomenology of the scalar mesons below $2\gev$~.

Summarizing, the present information from phenomenology does not give an unambiguous estimate for $\Lambda_{G}$. 
In this respect, an independent determination of $\Lambda_{G}$, such as the
one that we shall describe in the following, can be very useful in order to estimate the width of the scalar glueball and, along with it, the possibility of its experimental discovery.

\section{Tree-level scattering}
\label{sec:tree}
We expand the potential of Eq.~\eqref{potential} in powers of $G$:
\begin{align}
V(G)  &=-\frac{1}{16} \Lambda_{G}^4+\frac{1}{2}m_{G}^{2}G^{2}+\frac{1}{3!}\left(
5\frac{m_{G}^{2}}{\Lambda_{G}}\right)  G^{3}+\frac{1}{4!}\left(  11\frac
{m_{G}^{2}}{\Lambda_{G}^{2}}\right)  G^{4}\nonumber\\
& \quad +\frac{1}{5!}\left(  6\frac{m_{G}^{2}}{\Lambda_{G}^{3}}\right)  G^{5}%
+\frac{1}{6!}\left(  -6\frac{m_{G}^{2}}{\Lambda_{G}^{4}}\right)  G^{6}+...
\label{TaylorScattTheory:1972}%
\end{align}

\noindent Kinematics and conventions are summerized in Appendix~\ref{app:kinematics}. The tree-level $GG\rightarrow GG$ scattering amplitude can
be easily obtained from the $G^{3}$ and $G^{4}$ contributions as:%
\begin{equation}
A(s,t,u)=-11\frac{m_{G}^{2}}{\Lambda_{G}^{2}}-\left(  5\frac{m_{G}^{2}%
}{\Lambda_{G}}\right)  ^{2}\frac{1}{s-m_{G}^{2}}-\left(  5\frac{m_{G}^{2}%
}{\Lambda_{G}}\right)  ^{2}\frac{1}{t-m_{G}^{2}}-\left(  5\frac{m_{G}^{2}%
}{\Lambda_{G}}\right)  ^{2}\frac{1}{u-m_{G}^{2}} \text{ .}\label{totampl}%
\end{equation}
Due to the behaviors $m_G \propto N_c^{0}$ and $\Lambda_G \propto N_c$, it follows that the amplitude $A(s,t,u)$ scales as $N_c^{-2}$ as expected.

\begin{figure}[t]
\includegraphics[height=.21\textwidth]{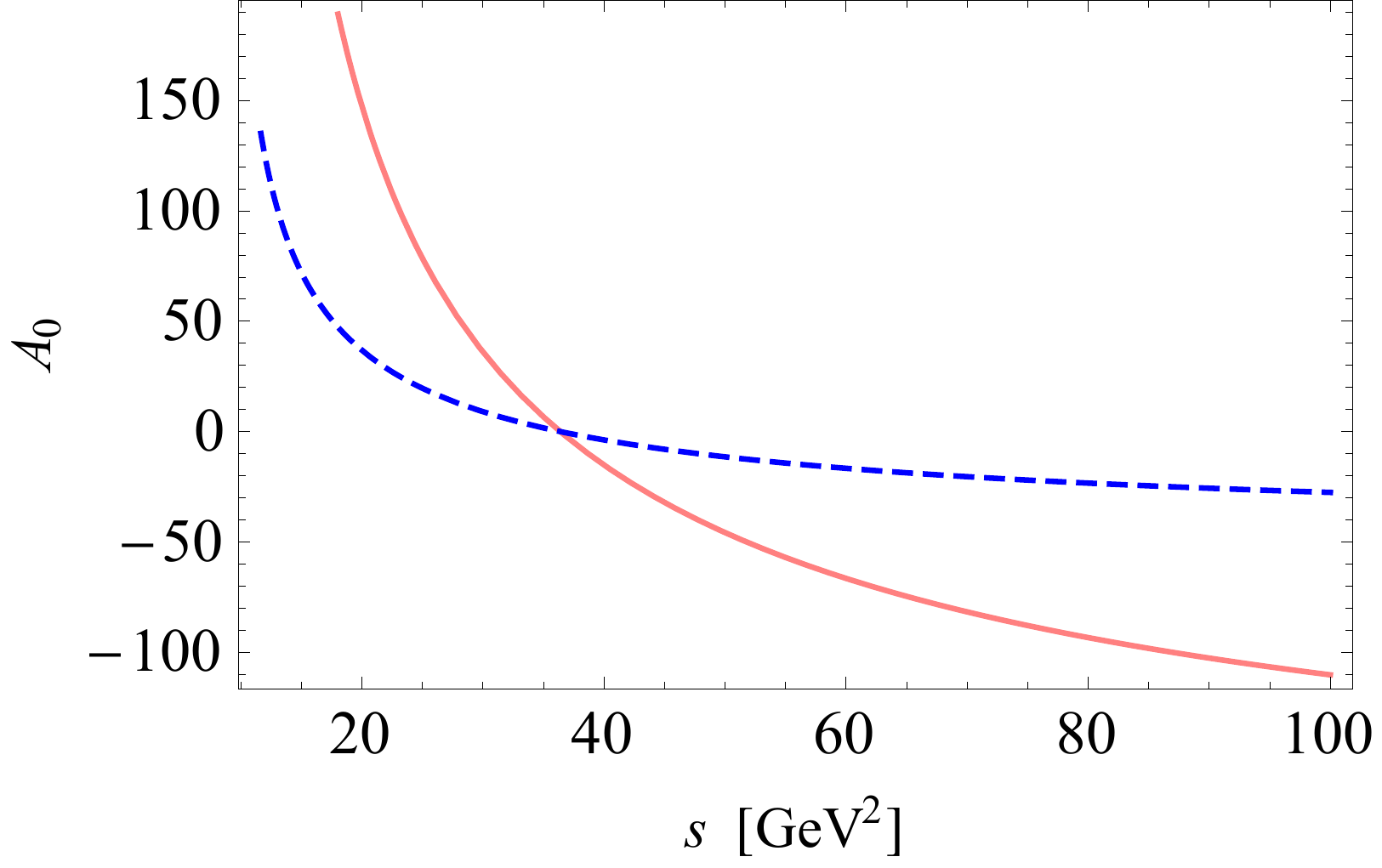}
\includegraphics[height=.21\textwidth]{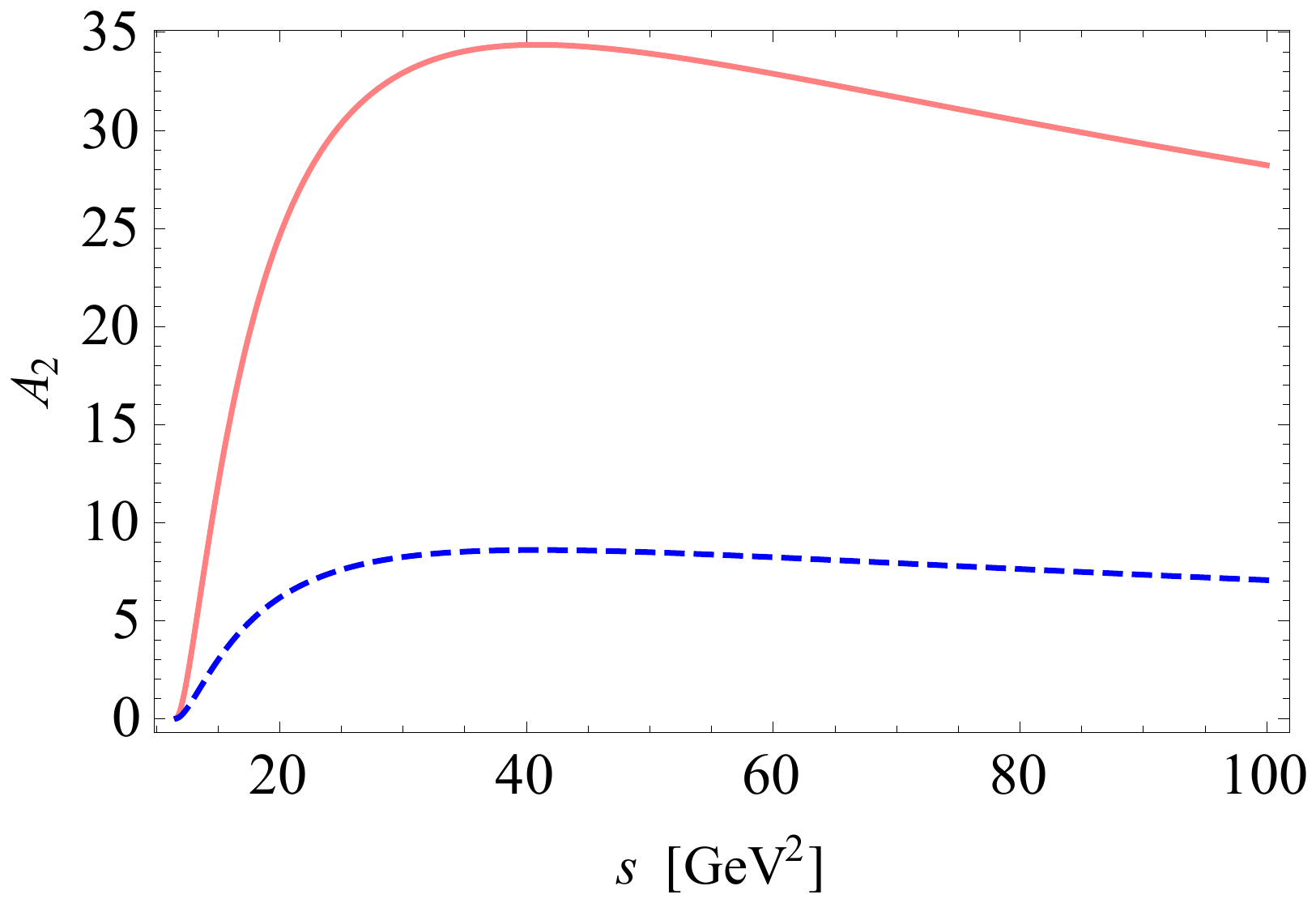}
\includegraphics[height=.21\textwidth]{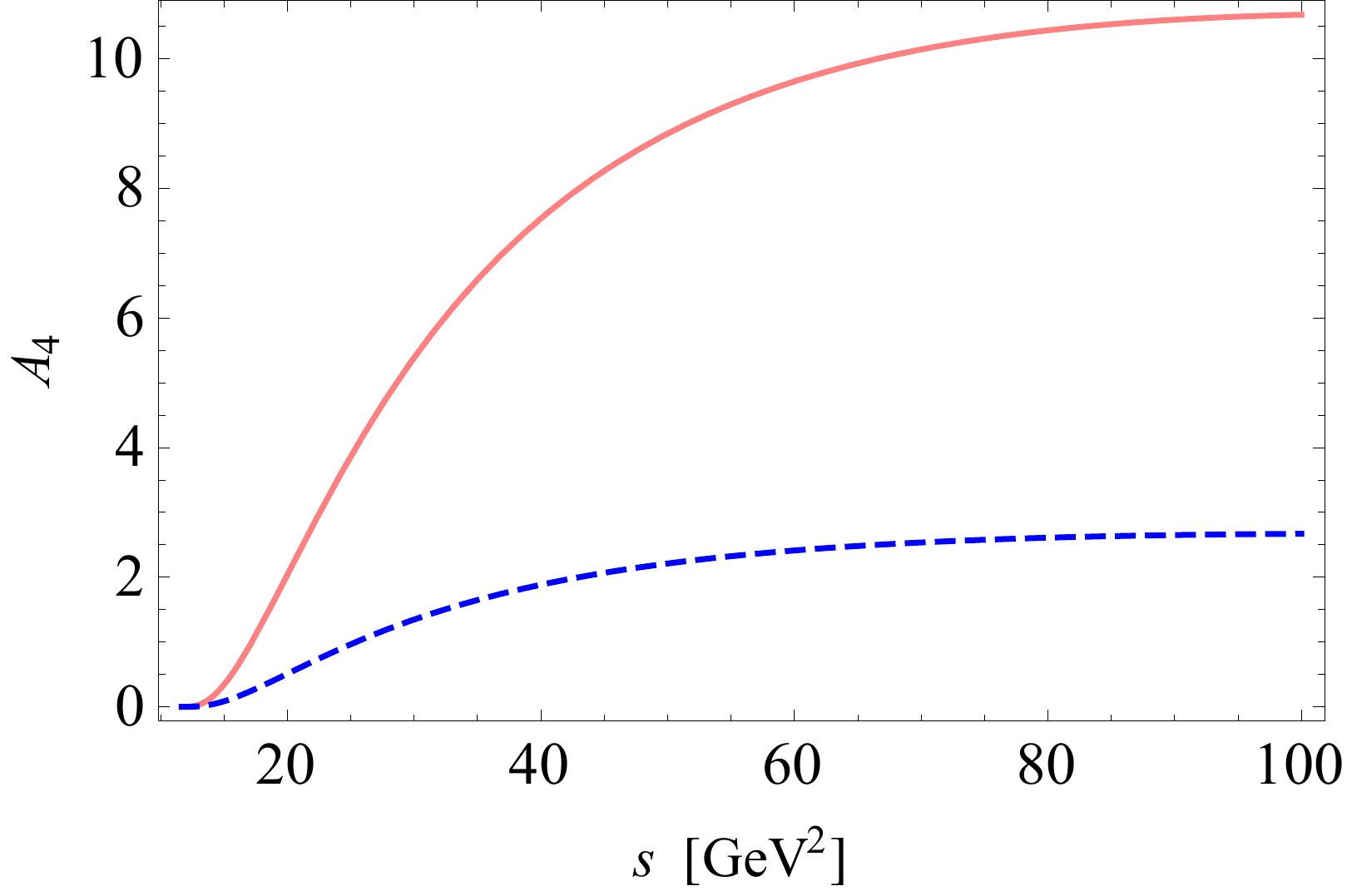}
\caption{The tree-level amplitudes $S$-, $D$-, and $G$-wave for two values of $\Lambda_G$ ($0.4\gev$ (pink, solid) to $0.8\gev$ (blue, dash), respectively). We remember that the amplitudes are purely real at this order. }				\label{fig:1}
\end{figure}

Next, we turn to the three lowest non-vanishing partial waves. Their plot are shown in Fig.~\ref{fig:1}.

\subsection{$S$-wave}
\noindent The $S$-wave takes the explicit form:
\begin{equation}
A_{0}(s)= -11\frac{m_{G}^{2}}%
{\Lambda_{G}^{2}}-25\frac{m_{G}^{4}}{\Lambda_{G}^{2}}\frac{1}{s-m_{G}^{2}%
}+50\frac{m_{G}^{4}}{\Lambda_{G}^{2}}\frac{Q_0\!\left(1+\frac{m_G^2}{2k^2}\right)  }{2k^2}
=-11\frac{m_{G}^{2}}%
{\Lambda_{G}^{2}}-25\frac{m_{G}^{4}}{\Lambda_{G}^{2}}\frac{1}{s-m_{G}^{2}%
}+50\frac{m_{G}^{4}}{\Lambda_{G}^{2}}\frac{\log\left(  1+\frac{s-4m_{G}^{2}%
}{m_{G}^{2}}\right)  }{s-4m_{G}^{2}}\,, \label{eq:A0}%
\end{equation}
where $Q_\ell(z')$ is the second-kind Legendre function. Close to threshold, $A_0(s)$ can be approximated by:
\begin{equation}
A_{0}(s)\simeq\frac{92m_{G}^{2}}{3\Lambda_{G}^{2}}-\frac{800}{9\Lambda_{G}%
^{2}}k^2+\dots
\end{equation}

Two general features of $A_{0}(s)$ are noteworthy:
$(i)$ there is a pole  for $s=m_{G}^{2}$ that corresponds to the propagation of a
single glueball in the $s$-channel; $(ii)$ the amplitude is also singular at $s=3m_{G}^{2}$, which is caused by the left-hand cut
generated by the projection onto the $S$-wave of the $t$- and the $u$-channel single pole~\cite{Frazer:1969euo,TaylorScattTheory:1972}.

The scattering length can be calculated as: 
\begin{align}
a_{0}^{\text{tree}}  &  = \frac{1}{32\pi m_{G}}\frac{92m_{G}^{2}}{3\Lambda_{G}^{2}}=\frac{23m_{G}%
}{24\pi\Lambda_{G}^{2}}\,.
\end{align}
Note that, since $\Lambda_G \propto N_c$, $a_{0}^{\text{tree}}\propto N_{c}^{-2}$ ,
in agreement with the expectations for the glueball-glueball scattering
amplitude~\cite{tHooft:1973alw}.

It is interesting to notice that there is a certain value $s_{c}$
for which the tree-level $S$-wave amplitude vanishes and then the phase-shift is a
multiple of $\pi$.
At this particular energy there is no interaction between the two
scattering glueballs, and consequently the interaction is soft in this neighborhood.
At tree level the value of $s_{c}$ is independent of $\Lambda_G$:
\begin{equation}
s_{c}\simeq12.59m_{G}^{2}\label{scb}.
\end{equation}
The vanishing of the scattering amplitude is valid as long as we can neglect
heavier glueballs, see next section. Nevertheless, the existence of a zero in the amplitude
offers an important check of the validity of our approach: the unitarization
procedure of Sec.~\ref{sec:unitarization} does not change the value of
$s_{c}$ and the behavior of the amplitude in its neighborhood. Namely, when the amplitude is small, it agrees with its unitarized counterpart in Eq.~\eqref{eqaluni}.

\subsection{$D$-wave}

For $\ell = 2$ we obtain
\begin{equation}
A_{2}(s)= 50 \frac{m_G^4}{\Lambda_G^2} \frac{Q_2\!\left(1+\frac{m_G^2}{2k^2}\right)}{2k^2}= \frac{50m_{G}^{4}}{(s-4m_{G}^{2})^{3}\Lambda_{G}^{2}}\left[
-3\left(  8m_{G}^{4}-6m_{G}^{2}s+s^{2}\right)  +\left[ s^2 -  2m_{G}^{4} -2m_{G}%
^{2}s\right)  \log\left(  1+\frac{s-4m_{G}^{2}}{m_{G}^{2}}\right)
\right] .
\end{equation}
Close to threshold, it reads:
\begin{equation}
A_{2}(s)\simeq\frac{5(s-4m_{G}^{2})^{2}}{3m_{G}^{2}\Lambda_{G}^{2}}+\cdots \text{
}=\frac{80k^{4}}{3m_{G}^{2}\Lambda_{G}^{2}}+\cdots ,%
\end{equation}
leading to a scattering length
\begin{equation}
a_{2}^{\text{tree}}=\frac{5}{6\pi m_{G}^{3}\Lambda_{G}^{2}}\,,
\end{equation}
to compare with the generic formula in Appendix~\ref{app:length}.

\subsection{$G$-wave}

For $\ell = 4$ we obtain:
\begin{align}
A_{4}(s) &= 50 \frac{m_G^4}{\Lambda_G^2} \frac{Q_4\!\left(1+\frac{m_G^2}{2k^2}\right)}{2k^2} =\frac{-25m_{G}^{4}}{3(s-4m_{G}^{2})^{5}\Lambda_{G}^{2}%
}\Big[-5\left(  46m_{G}^{4}-2m_{G}^{2}s-5s^{2}\right)  \left(  s-2m_{G}%
^{2}\right)  \left(  s-4m_{G}^{2}\right)  \nonumber\\
&\quad  +6\left(  74m_{G}^{8}-124m_{G}^{6}s+54m_{G}^{4}s^{2}-4m_{G}^{2}s^{3}%
-s^{4}\right)  \log\left(  1+\frac{s-4m_{G}^{2}}{m_{G}^{2}}\right)  \Big].
\end{align}
For $s$ close to threshold, the quantity $A_{4}(s)$ is approximated by:%
\begin{equation}
A_{4}(s)\simeq\frac{5(s-4m_{G}^{2})^{4}}{63m_{G}^{6}\Lambda_{G}^{2}}+\cdots = \frac{5\cdot 4^4 k^8}{63m_{G}^{6}\Lambda_{G}^{2}}+\cdots\,,
\end{equation}
leading to the scattering length:
\begin{equation}
a_{4}^{\text{tree}}=\frac{40}{63\pi m_{G}^{7}\Lambda_{G}^{2}}.
\end{equation}

\section{Unitarization} 
\label{sec:unitarization}
\begin{figure}[t]
\centering
\includegraphics[width=0.5\textwidth]{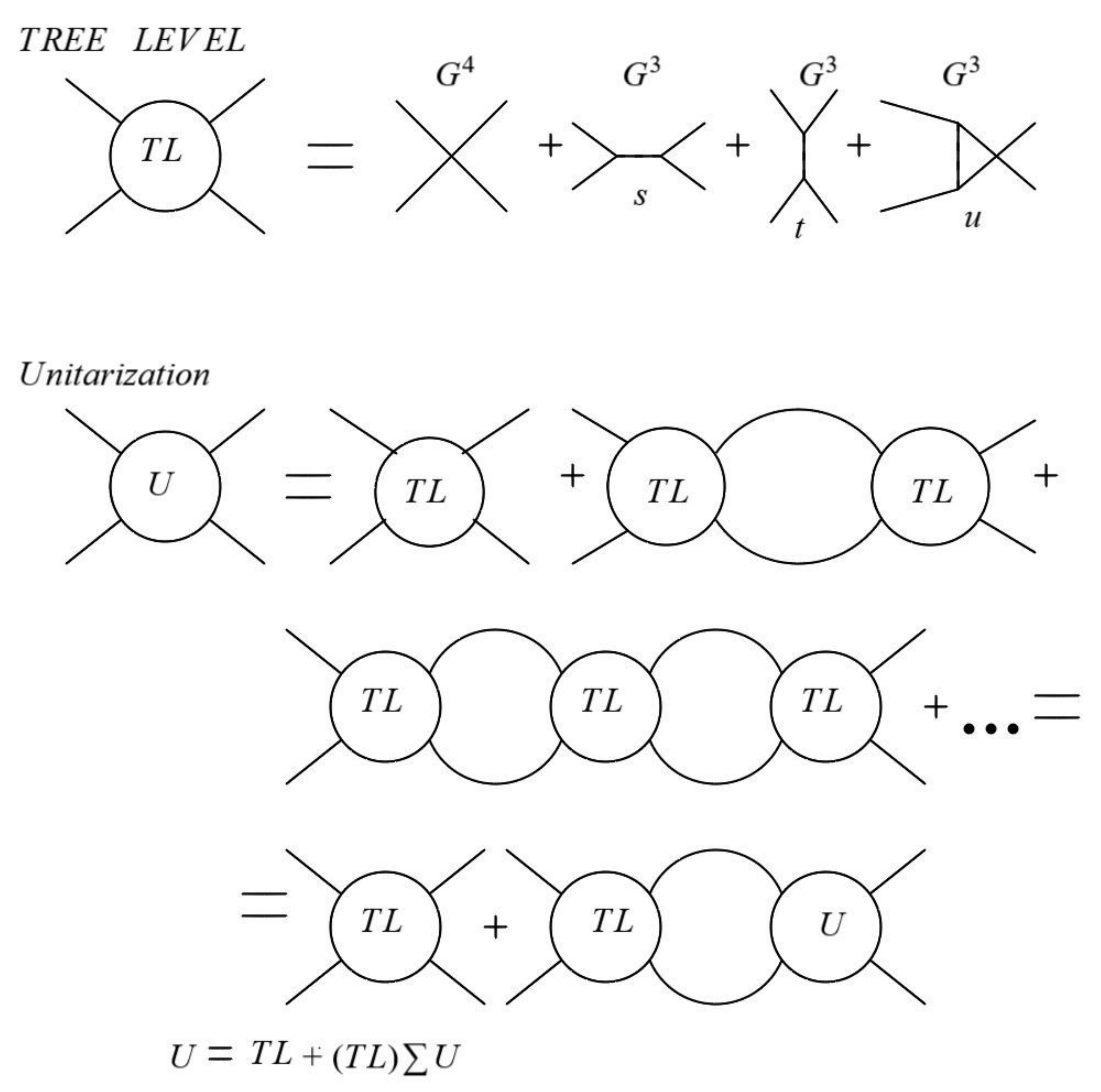}
\caption{Upper part: tree-level scattering diagrams. \\  
Lower part: schematic representation of the unitarization.}
\label{fig:2}
\end{figure}

It is well known that fixed-order calculations do not provide a reasonable approximation to the amplitude at finite distance from threshold. In particular, they cannot produce (bound state) poles, unless they are hardcoded in the lagrangian.
To this end, one must unitarize the partial wave amplitudes. Such a procedure is non unique, and several schemes have been proposed, in particular for chiral lagrangians~\cite{Truong:1991gv,GomezNicola:2001as,Selyugin:2007jf,Cudell:2008yb,Nebreda:2011di,Delgado:2015kxa,Oller:2020guq}.
Here we adopt a scheme also known as ``on-shell approximation'' \cite{Dobado:1992ha,Oller:1997ng}:
\begin{equation}
A_{\ell}^{\text{U}}(s)=\left[  A_{\ell}^{-1}(s)-\Sigma(s)\right]  ^{-1}= \dfrac{A_{\ell}(s)}{1-A_{\ell}(s)\Sigma(s)} ,
\label{eqaluni}%
\end{equation}
where $\Sigma(s)$ is a glueball-glueball self-energy loop function, whose
imaginary part is fixed by:
\begin{equation}
\im\Sigma(s)=\theta\!\left(s - 4m_G^2\right)
\frac{1}{2}\frac{1}{16\pi}\sqrt{1-\frac{4m_{G}^{2}}{s}}=\theta\!\left(s - 4m_G^2\right)\frac{1}{2}\frac
{k}{8\pi\sqrt{s}} \label{imloop} \text{ ,}%
\end{equation}
\ie to the relativistic 2-body phase space of two identical particles.
Clearly, when $A_{\ell}(s)$ is sufficiently small, $A_{\ell}^{\text{U}}(s)\simeq
A_{\ell}(s)$. The diagrammatic representation of the resummation is presented in
Fig.~\ref{fig:2}.

\begin{figure}[t]
\centering
\includegraphics[width=0.5\textwidth]{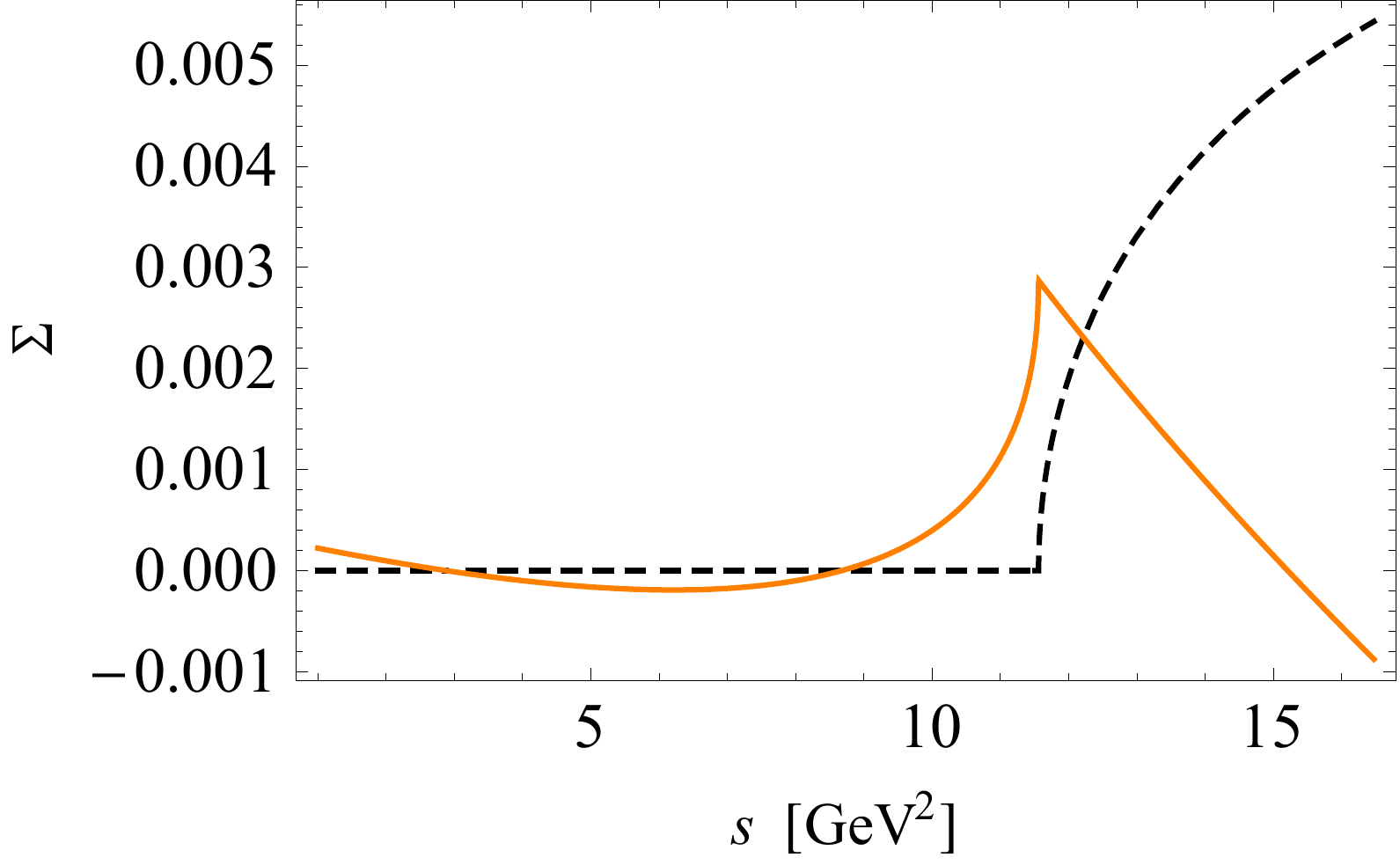}
\caption{Loop function $\Sigma(s)$ of Eq.~\eqref{loop}, real (orange, solid)  and imaginary part (black, dash). The subtractions are chosen to make $\Sigma(s)$ vanish at $s=m_G^2$ and~$3m_G^2$. By construction, the imaginary part is nonzero above $s > 4m_G^2$.}
\label{fig:3}
\end{figure}

Since the analytic properties of $\Sigma(s)$ are known, one can use dispersion relations to reconstruct the real part from its imaginary part, up to a polynomial. Here, we fix the latter by subtracting $\Sigma(s)$ twice, such that $(i)$~we preserve the single-glueball pole at $s=m_{G}^{2}$ (in other words, the
tree-level mass is preserved also at the unitarized level), and $(ii)$ we require that the unitarized amplitude coincides 
with the tree-level one in the neighborhood of the left-hand branch point. Although this prescription is ad-hoc, it attempts at preserving the cross-channel poles upon resummation of the partial waves. Both features can be obtained by requiring:
\begin{equation}
\Sigma(s=m_{G}^{2})=0\text{ and }\Sigma(s=3m_{G}^{2})=0\text{ .}
\label{constraints}%
\end{equation}
Then, close to $m_{G}^{2}$ and to $3m_{G}^{2}$ it also holds that
$A_{\ell}^{\text{U}}(s)\simeq A_{\ell}(s).$ The needed form of the loop function is
given by:

\begin{equation}
\Sigma(s)=\dfrac{(s-m_{G}^{2})(s-3m_{G}^{2})}{\pi}\int_{4m_G^{2}}^{\infty}%
\frac{\im \Sigma(s^{\prime})}{(s^{\prime}-s-i\varepsilon)(s^{\prime}-m_{G}%
^{2})(s^{\prime}-3m_{G}^{2})}ds^{\prime}\text{ .} \label{loop}%
\end{equation}
The loop function is plotted in Fig.~\ref{fig:3}: the imaginary part is
expressed in Eq.~\eqref{imloop}, while the real part is evaluated from the
previous equation; the principal value prescription is understood for $s>4m_{G}^{2}$.
For our purposes, the $S$-wave amplitude is the most important
one. The two subtractions preserve the direct-channel single-glueball pole,
as well as the behavior of the amplitude in the vicinity of the logarithmic branch point due to the single-glueball exchange in the $t$- and
$u$-channels. Note, while the convergence of the loop function $\Sigma(s)$ is
guaranteed by a single subtraction, some unphysical features (such as ghost
poles) would emerge, see the comments at the end of this Section.

\noindent
Next, we turn to the unitarized scattering length. At threshold, $\Sigma(4m_{G}%
^{2})=(64\pi\sqrt{3})^{-1}\simeq 0.028715$, which modifies the scattering length as:
\begin{equation}
a_{0}^{\text{U}}=\frac{1}{32\pi m_{G}}A_{0}^{\text{U}}(s)=\frac{1}{32\pi
m_{G}}\frac{1}{\frac{3\Lambda_{G}^{2}}{92m_{G}^{2}}-\Sigma(4m_{G}^2)}.
\end{equation}
The scattering length diverges when $\frac{3\Lambda_{G}^{2}}{92m_{G}^{2}} =\Sigma(4m_{G}^2)$. Numerically, the critical value for
$\Lambda_{G}$ reads:

\begin{equation}
\Lambda_{G,\text{crit}}=m_{G}\sqrt{\frac{92}{3}\cdot \frac{1}{64\pi\sqrt{3}}} \simeq 0.2967m_{G}%
\sim 0.504\gev,%
\end{equation}
where on the right hand side $m_{G}=1.7\gev$ was used. The divergence of the scattering
length points to the emergence of a glueball-glueball bound state. One can
therefore summarize the situation as follows:
\begin{align}
\text{for }\Lambda_{G}  >\Lambda_{G,\text{crit}} &\text{: no bound state and
}a_{0}^{\text{U}}>0\nonumber \text{ ,}\\
\text{for }\Lambda_{G} \to \Lambda_{G,\text{crit}} &\text{: glueballonium at
threshold and~} a_0^\text{U} = \infty\nonumber \text{ ,}\\
\text{for }\Lambda_{G} <\Lambda_{G,\text{crit}} &\text{: glueballonium
with } m_{B}\in(\sqrt{3}m_{G},2m_{G}),\,a_{0}^{\text{U}}<0 \text{ .}
\end{align}

\begin{figure}[t]
\centering
\includegraphics[width=0.5\textwidth]{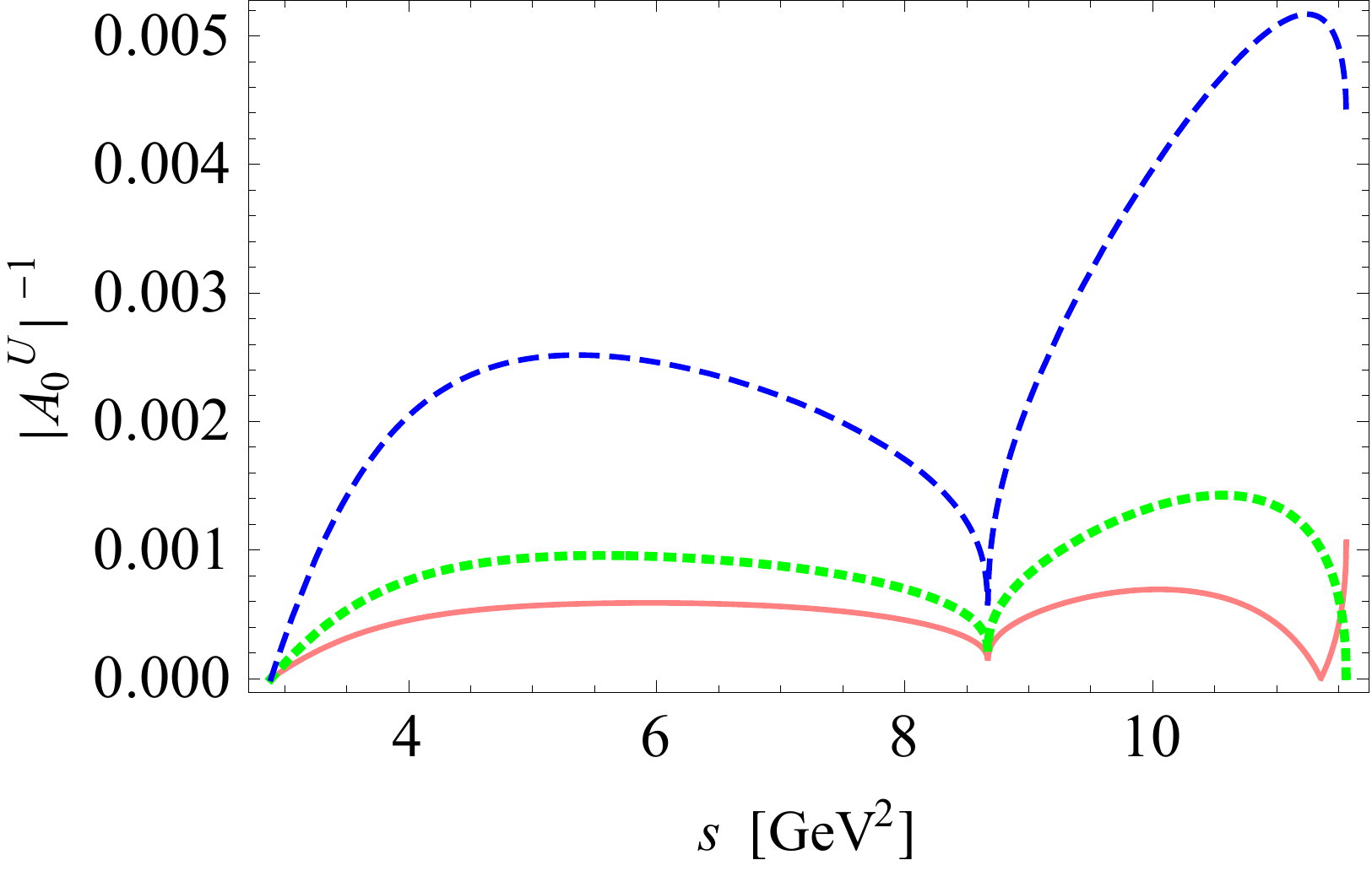}
\caption{The inverse amplitude $|A^U_0|^{-1}$ as function of $s$ for  $\Lambda_G=0.4\gev$ (pink, solid), $\Lambda_G = \Lambda_{G,\text{crit}} \approx 0.504\gev$ (green, dot), and $\Lambda_G=0.8\gev$ (blue, dash), text for details. }
\label{fig:4}
\end{figure}

\begin{figure}[t]
\centering
\includegraphics[width=0.5\textwidth]{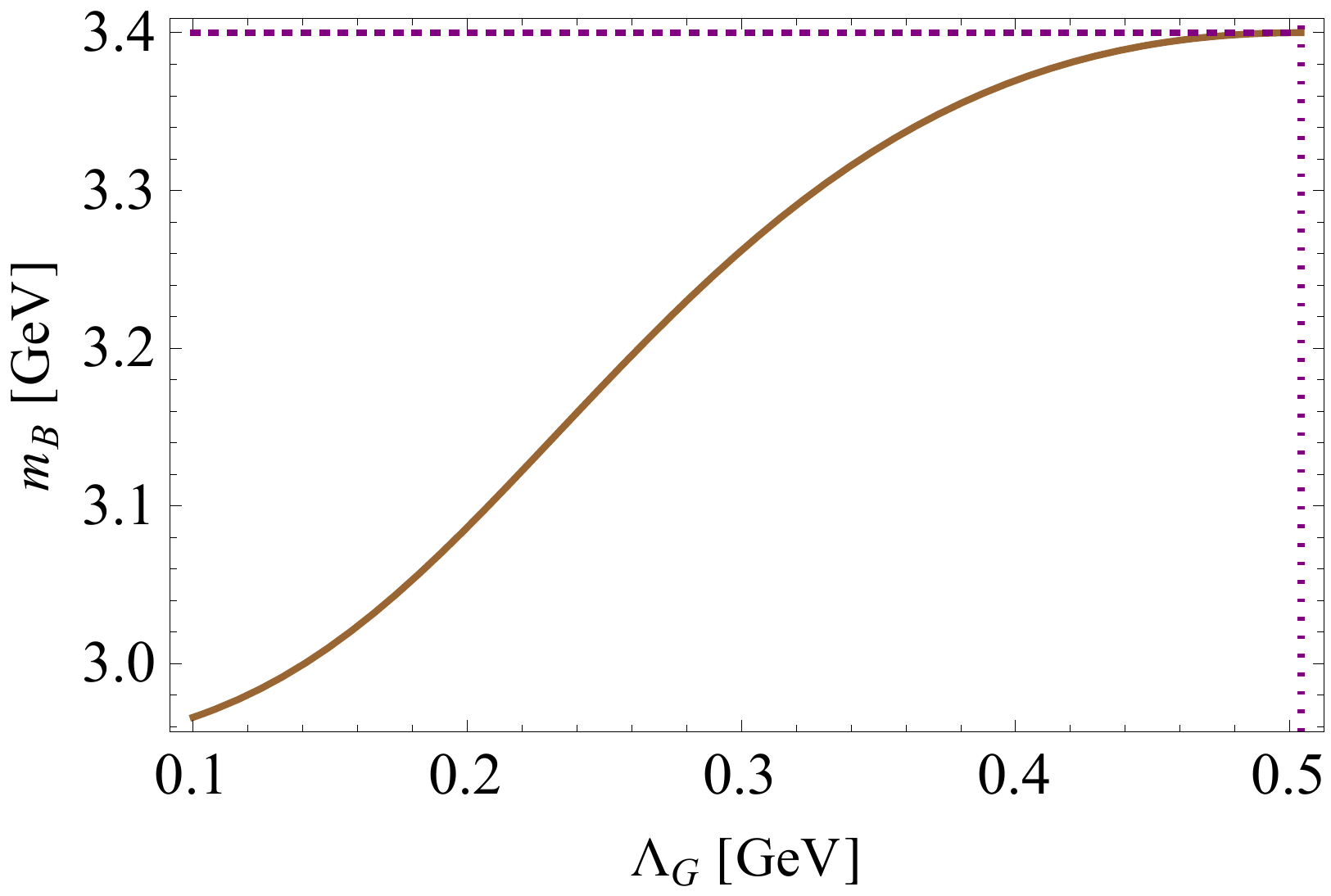}
\caption{The mass of the glueballonium as function of the parameter $\Lambda_{G}$.}
\label{fig:5}
\end{figure}

\noindent
The inverse amplitude is plotted in Fig.~\ref{fig:4} for three choices of
$\Lambda_{G}$. The zero at $m_{G}^{2}$
(single particle pole) and $s=3m_{G}^{2}$ (branch point) is always present. For $\Lambda
_{G} > \Lambda_{G,\text{crit}}$ no other zero is present: the attraction is not strong enough to
generate any bound state. For $\Lambda_{G} \rightarrow \Lambda_{G,\text{crit}} \simeq 0.504\gev$ a third zero
appears at threshold, which corresponds to the 
glueballonium pole. For smaller $\Lambda_{G}$ this zero moves to lower
value of $s$.
The mass of the glueballonium $m_{B}$ can thus be found as function of $\Lambda_{G}$ as a solution of the equation:
\begin{equation}
A_{0}^{-1}(s,m_{G},\Lambda_{G})-\Sigma(s,m_{G})=0\text{ for }s\in(3m_{G}%
^{2},4m_{G}^{2})\text{ ,}%
\end{equation}
which is also plotted in Fig.~\ref{fig:5}.
As expected, $\Lambda_{G}$ descreases with the bound
state mass. We recall that lattice QCD estimates $\Lambda_{G}=0.4\gev$,
which implies $m_{B}=3.37\gev$. Moreover, in the limit $\Lambda_{G}\rightarrow0$ (that is,
when the coupling constant tends to infinity) the mass of the glueballonium
tends to $\sqrt{3}m_{G}.$ 
Yet, this is a feature of the adopted unitarization scheme that does not hold in general, see Appendix~\ref{app:noverd}. 
For the
practical case of the glueballonium, we regard $\Lambda_{G}\simeq0.3\gev$  as a
lower limit for this parameter, thus the corresponding mass of the
glueballonium is still quite close to threshold where the influence of the
unitarization scheme is expected to be less prominent.

According to the results obtained within our unitarization scheme,
the glueballonium is an additional scalar state that could be identified with the excited glueball at $\sim 3\gev$ found in~\cite{Morningstar:1999rf}.
Eventually it could be found in
experiments as yet another additional meson
that does not fit in the quark-antiquark picture,
if it and the constituent glueballs are not too broad. 
At present the decay width of the lightest
glueball is unknown.
Hopefully, this work may provide an estimate for it, 
since it depends on $\Lambda_{G}$.
Similarly, also the
width of the glueballonium could be determined by studying its
two- and four-body decays.

\begin{figure}[t]
\includegraphics[height=.2\textwidth]{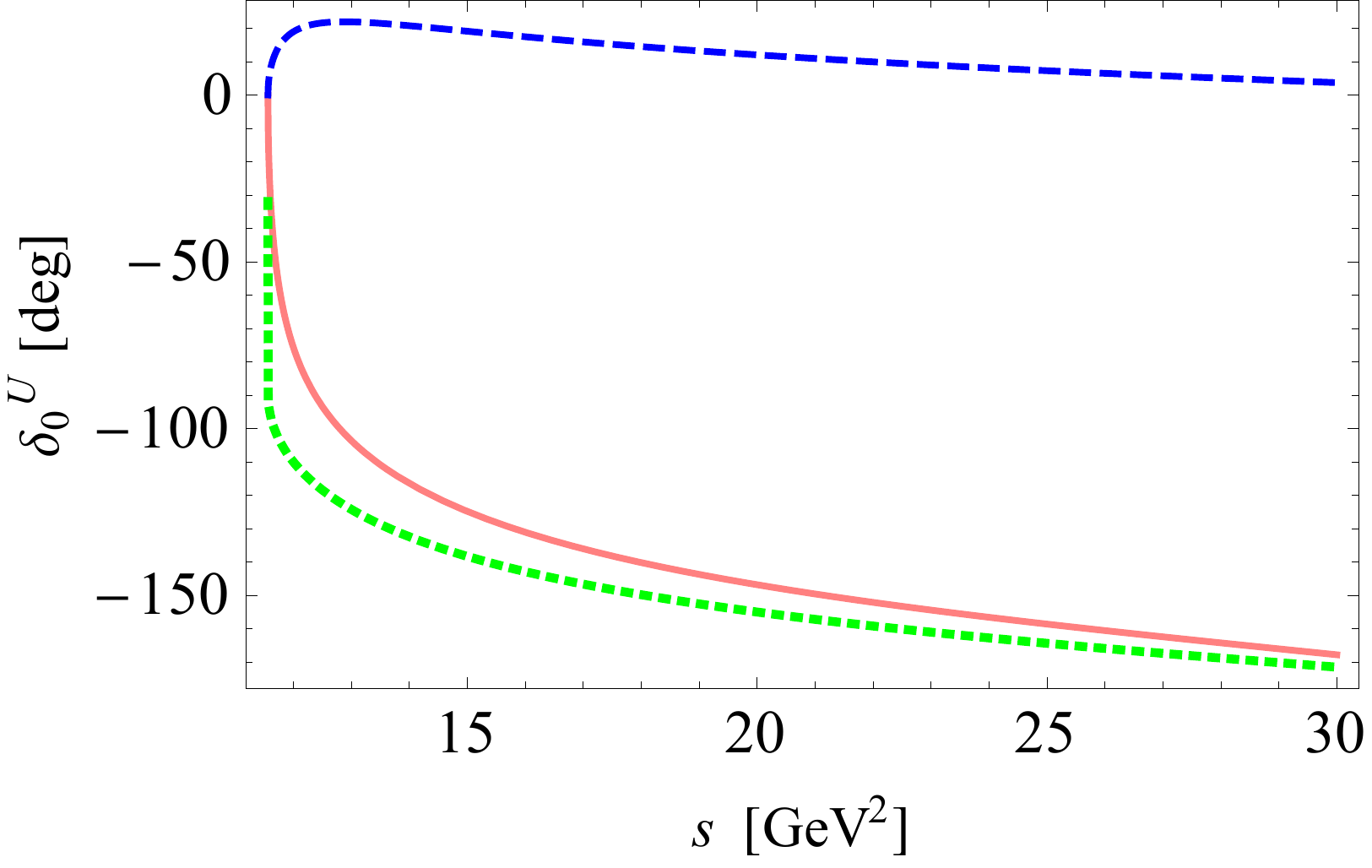}
\includegraphics[height=.2\textwidth]{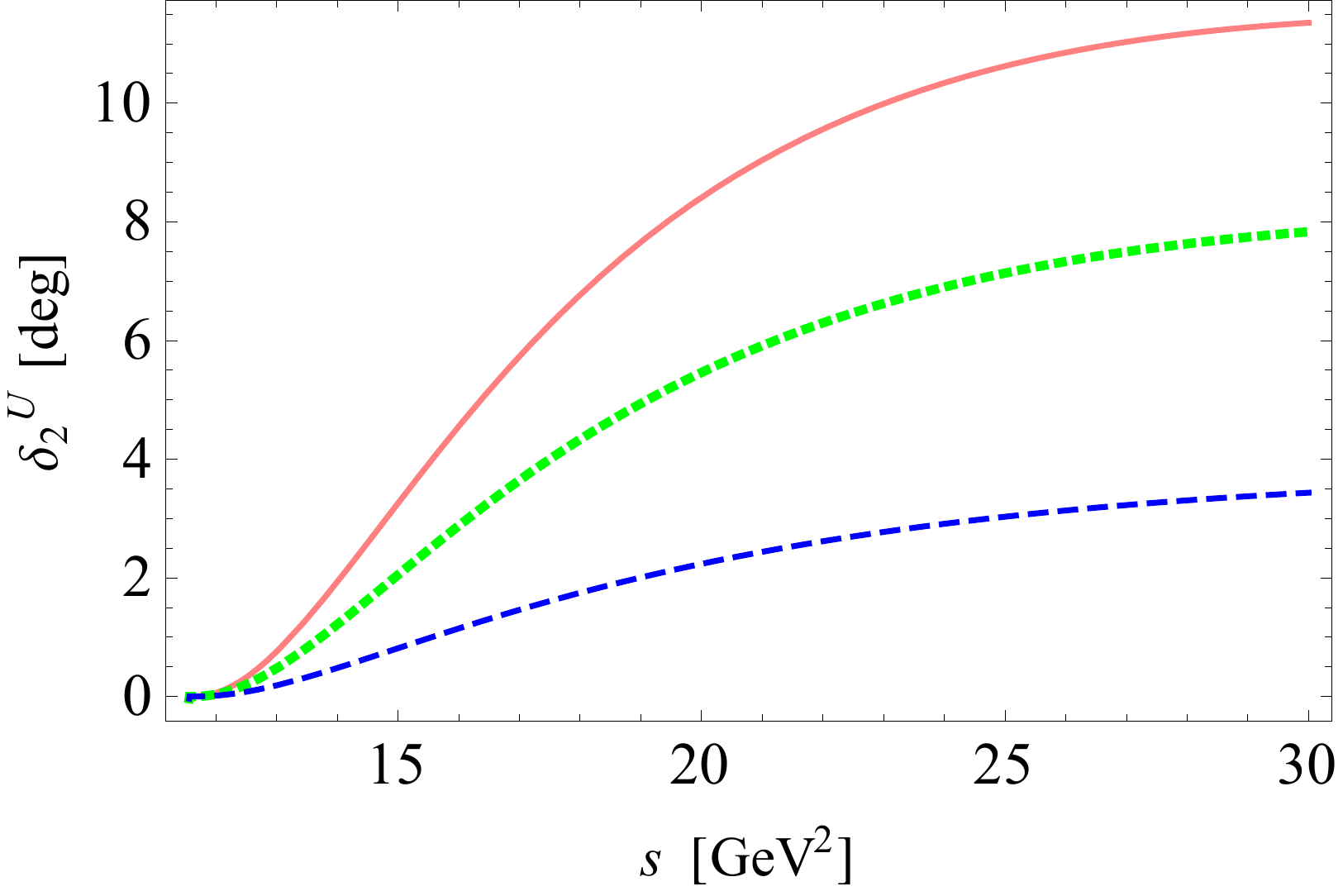}
\includegraphics[height=.2\textwidth]{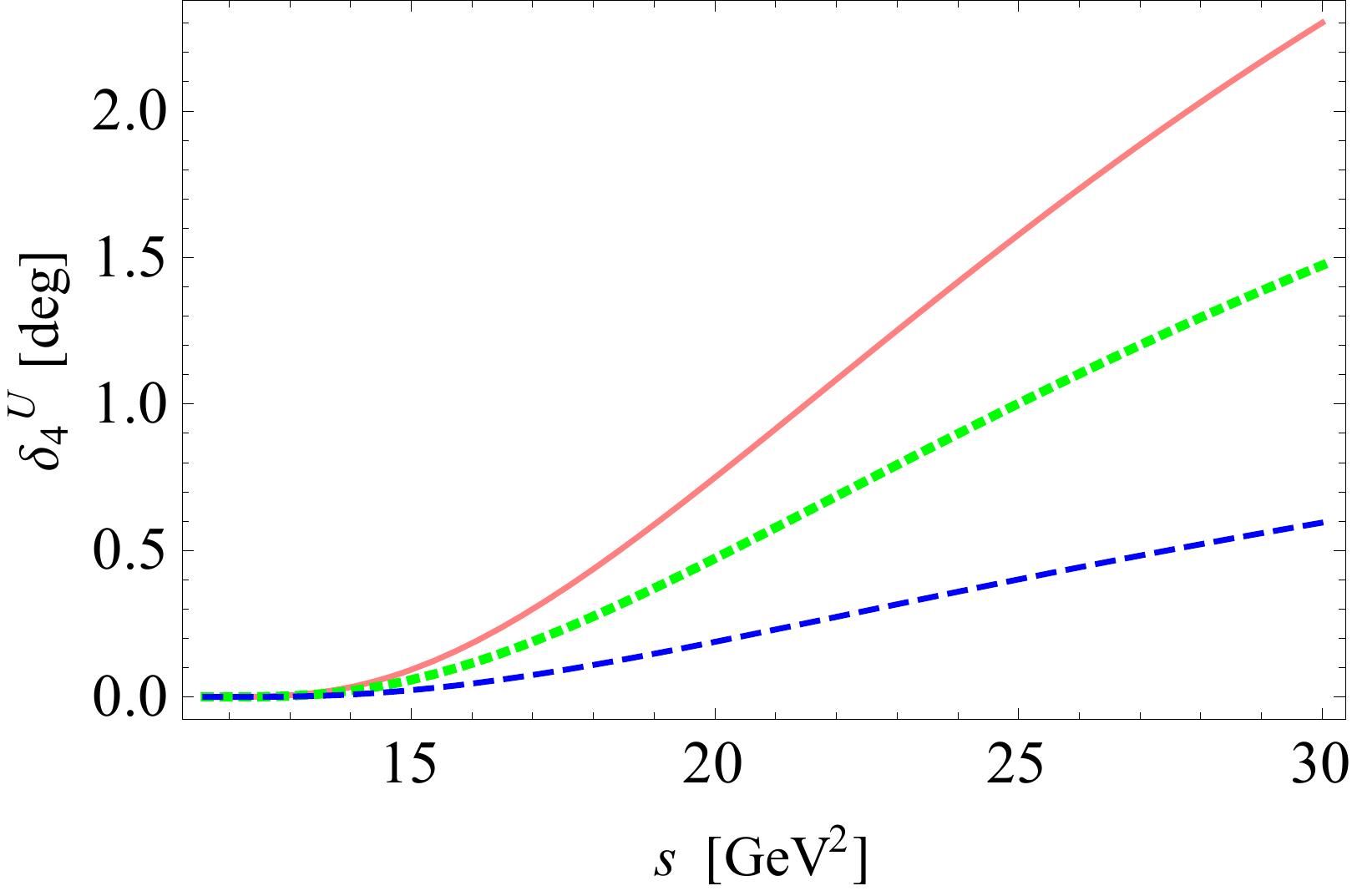}
\caption{The phase shift of the unitarized amplitude for $S$- (left), $D$- (center), and $G$-wave (right) for the values $\Lambda_G=0.4\gev$ (pink, solid), $0.5041\gev$ (green, dot), and $0.8\gev$ (blue, dash). The functions are plotted for values of $s$ only slightly higher than the elastic window. \label{fig:6}} 
\end{figure}

\noindent
In Fig.~\ref{fig:6} the unitarized phase shifts are shown for different values of $\Lambda
_{G}$. Higher waves have small phase motion:
at threshold they increase as $k^{2\ell +1}$, reach a maximum of few degrees, and slowly approach zero for large
$s.$\\
The behavior of the $S$-wave is 
more interesting: for $\Lambda_{G}$ above the critical value, the phase shift reaches a maximum, then decreases and vanishes at $s=s_{c}$ defined in Eq.~\eqref{scb}. 
After this value, the phase shift becomes negative and approaches
$-180^{\circ}$ for large values of $s$.
For $\Lambda_{G} \le \Lambda_{G,\text{crit}}$ the behavior of
the phase shift is utterly different: it starts negative already at threshold and, as a consequence of the existence of the glueballonium, it
reaches $-180^{\circ}$ at $s_{c}$ and further decreases  to $-360^{\circ}$
asymptotically.
The amplitudes are plotted up to values of $s$ only slightly higher than the elastic window $(3m_{G})^{2}\simeq26\gevsq$. To appreciate the asymptotic behavior, we refer to Appendix~\ref{app:asymptotics} for plots of
larger values of $s$.

As anticipated, the asymptotic behavior of $\delta_{0}^{\text{U}}(s)$ is explained by Levinson's theorem~\cite{TaylorScattTheory:1972,Hartle:1965nj,Ma:2006zzc}: according to
which the number $n$ of poles below threshold (including bound states) is
related to the phase difference
$\delta_{0}^{U}(s_{th})-\delta_{0}^{U}(\infty)=n\pi$.
Since in our convention $\delta_{0}^{U}%
(s_{th})=0$,
\begin{subequations}
\begin{align}
\delta_{0}^{U}(\infty)  &  =-\pi\text{ for }\Lambda_{G}>\Lambda_{c}\text{
}\rightarrow n=1\rightarrow\text{one pole for }s=m_{G}^{2}\,;\\
\delta_{0}^{U}(\infty)  &  =-2\pi\text{ for }\Lambda_{G}\leq\Lambda_{c}\text{
}\rightarrow n=2\rightarrow\text{one pole for }s=m_{G}^{2}\text{ and one for
}s=m_{B}^{2}\,.%
\end{align}
\end{subequations}

\begin{figure}[t]
\includegraphics[height=.2\textwidth]{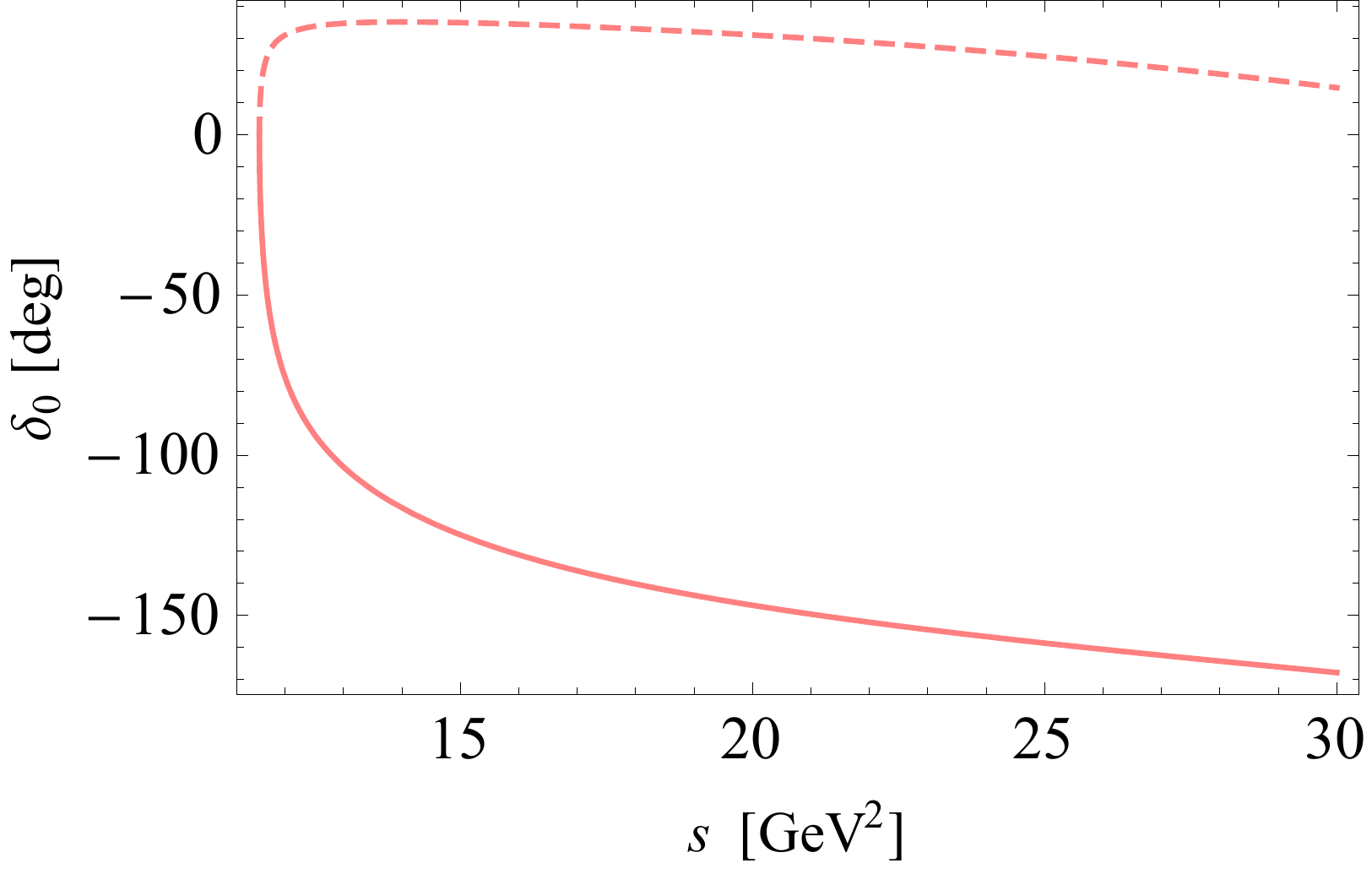}
\includegraphics[height=.2\textwidth]{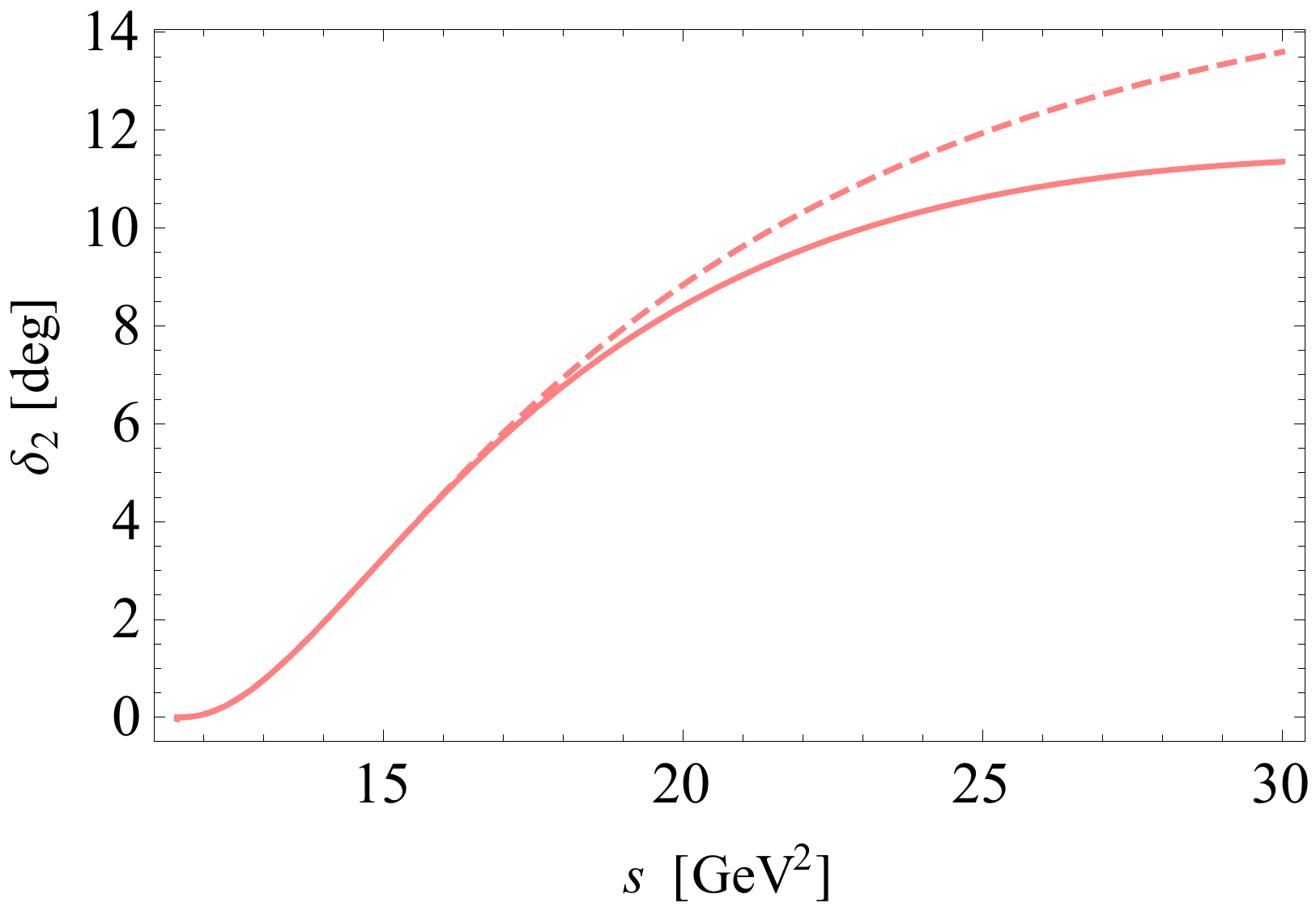}
\includegraphics[height=.2\textwidth]{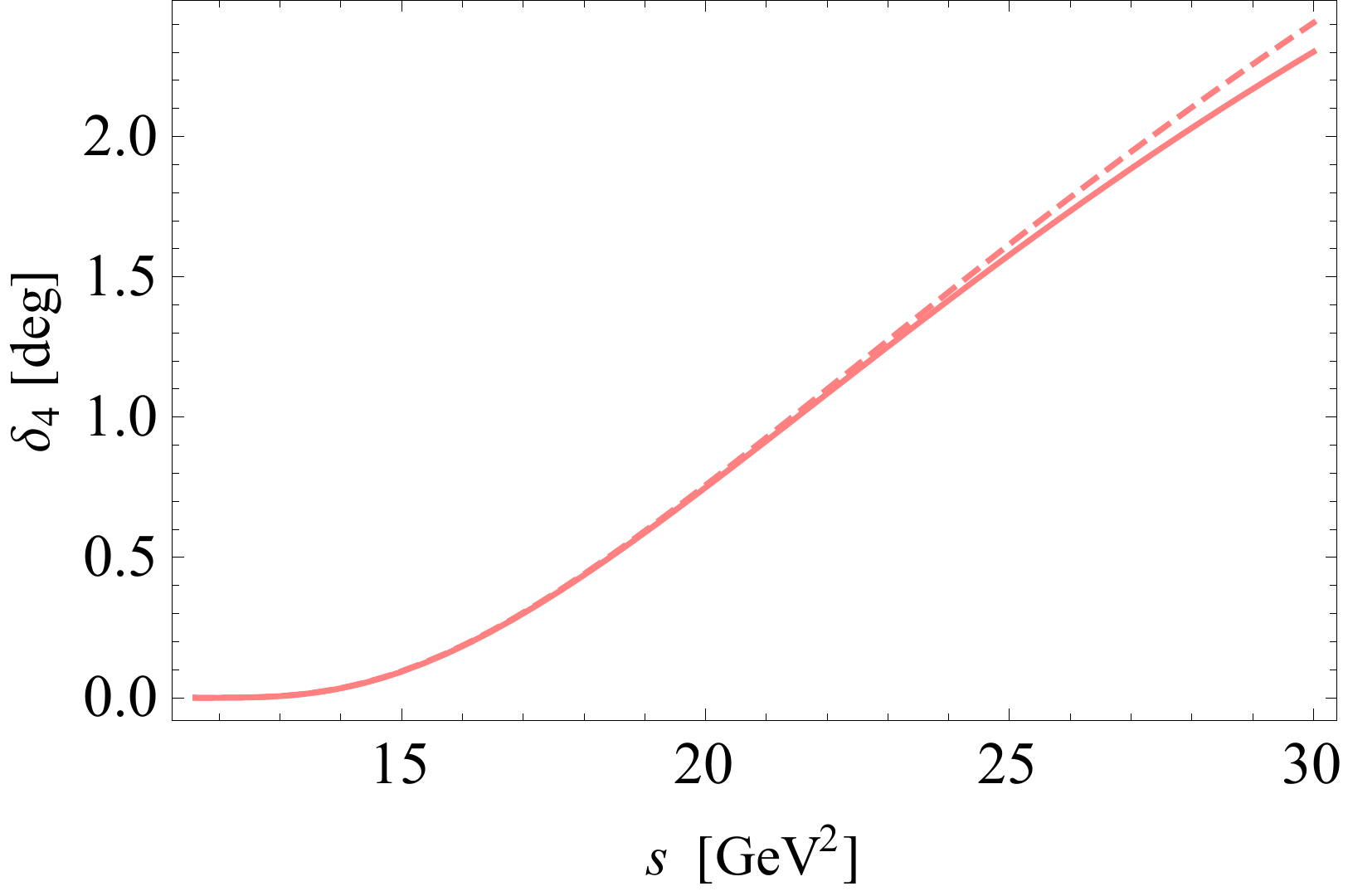}
\caption{Comparison of unitarized (solid) and  tree-level (dash) phase shifts as function of $s$ for $\Lambda_G=0.4\gev$. In the three panels we show the $S$-, $D$-, and $G$-waves.}				\label{fig:7}
\end{figure}

\noindent
For completeness, we compare in Fig.~\ref{fig:7} the phase shifts of the tree-level
and unitarized models. As
expected, higher waves are slightly affected by
unitarization, especially close to threshold. In particular, scattering lengths are unchanged. 
Conversely, the $S$-wave is highly affected by  unitarization, since a new bound state is created.

Finally, we conclude with some important remarks on the approximations implemented:
\begin{enumerate}
    \item The choice of the twice-subtracted loop of Eq.~\eqref{loop} fixes the single-particle pole and the branch point to the ones of the tree-level amplitude. By using,
instead, the once-subtracted loop that preserves the single-particle pole only would generate,
in addition to a glueballonium, a ghost with negative norm~\cite{Kummer:1962ay,Briscese:2018oyx,Donoghue:2019fcb}. For this reason, a single-subtraction cannot be used for the study of the emergence of a bound state when the potential involves both cubic and quartic interactions.
Yet, as recently shown in Ref.~\cite{Samanta:2020pez} in the case
of the quartic interaction only, the use of the loop with a single subtraction
generates a qualitative similar pattern for the bound state.
\item A unitarization procedure based on Lippmann-Schwinger equations would prescribe the potential to depend on the (off-shell) loop momentum. If the potential is approximated to be on-shell, it can be factored out of the integral, and one obtains the simple form of the scalar loop function. In this way, the unitarized amplitude is simply given by an algebraic equation and one does not need to solve the full integral equation.
Such simple unitarization schemes are quite common in the literature. The price to pay is a bad description of the left hand cut, and the possible emergence of unphysical singularities, as the ghost pole mentioned earlier. However, the effect of these drawbacks is generally milder close to threshold. 
\item In general, the results depend not only on the chosen subtraction, but also on
the unitarization scheme, see e.g. Refs.~\cite{Frazer:1969euo,Cavalcante:2004zm,Delgado:2015kxa,Selyugin:2007jf,Cudell:2008yb,Hayashi:1967bjx,Nebreda:2011di}. 
Yet, it is expected that the
overall phenomenology would be quite similar as long as the interaction is not too strong (i.e., the glueballonium is close enough to the threshold).  
A direct test using another unitarization scheme ---the so-called $N/D$ approach--- is reported in Appendix~\ref{app:noverd}. The results for the glueballonium turns out to be quite similar, as long as $\Lambda_G$ is larger than $0.3\gev$. In this case, the bound state mass is not far from threshold and the on-shell and $N/D$ schemes are quite close to each other. 
In the future, one should use other unitarization schemes as well as the higher order calculations in order to check further the stability of the results.
\item The imaginary part of the loop in Eq.~\eqref{imloop} is taken to be
valid up to indefinitely large values of $s$. 
This is not true, as multiparticle states will also contribute to the imaginary part.
Phenomenologically, this is often realized by introducing a smooth cutoff function, that can be interpreted as the overlap of the glueball wave functions.
Models for the latter have been discussed in Refs.~\cite{Giacosa:2004ug,Faessler:2003yf,Bowler:1994ir,Giacosa:2007bn}.
Yet, the form of this cutoff function is not known and would introduce a model
dependence of the results. It should be reconsidered when lattice data
will be available.
\end{enumerate}

\section{The (negligible) influence of other glueballs on $GG$ scattering}
\label{sec:heavy}
So far, we discussed $GG$ scattering within the dilaton potential, which contains a single scalar field
$G$. It is then natural to ask if other heavier glueballs can affect the
previous results, especially near the threshold $s_{th}=4m_{G}^{2}$ and in
connection to the formation of a bound state. In particular, a hypothetical
glueball of mass $\sim 3.5\gev$ which can be exchanged in the
$s$-channel of $GG$ scattering would eventually dominate the near-threshold
region, thus modifying our previous predictions. Indeed, as shown in Ref.~\cite{Chen:2005mg} there are various glueballs in that energy region. In the
following section, we shall show that under quite general assumptions relying
on dilatation invariance, no heavy glueball is exchanged in the $S$-channel,
thus not affecting the results discussed above.

\subsection{Heavy scalar glueball(s)}

We consider a heavy scalar glueball $H$, coupled to the ground state glueball. Since the original dilaton potential is $Z_2$ symmetric for the field $G$, we assume that this is the case also for the field $H$. The enlarged Lagrangian takes the form:
\begin{equation}%
\mathcal{L}=\mathcal{L}_\text{dil}+\frac{1}{2}(\partial_{\mu}H)^{2}-\frac{\alpha}{2}G^{2}H^{2}-\beta H^{4} \equiv \frac{1}{2}(\partial_{\mu}G)^{2} + \frac{1}{2}(\partial_{\mu}H)^{2} - V(G,H)
\end{equation}
where we have assumed that --besides $\mathcal{L}_\text{dil}$-- the other terms are dilatation invariant. 
Namely, the parameters $\alpha$ and $\beta$ are
dimensionless and the corresponding terms are dilatation invariant. A mass
term proportional to $H^{2}$ or three-leg interaction terms proportional to
$H^{2}G$ or $GH^{2}$ are excluded, since they corresponding coupling constants
are dimensionful and thus break dilatation invariance. This contradicts 
our assumption that only the logarithmic term for the dilaton
field $G$ is responsible for the trace anomaly.
The minimum of the potential is realized for
\begin{equation}
\left\{\begin{aligned}
\partial_{G}V(G,H) &  =\frac{m_{G}^{2}}{\Lambda_{G}^{2}}G^{3}\ln\frac
{G}{\Lambda_{G}}+\alpha GH^{2}=0\text{ ,}\\
\partial_{H}V(G,H) &  =\alpha G^{2}H + 4\beta
H^{3}=0\text{ ,}%
\end{aligned}\right.
\end{equation}
hence $G_{0}=\Lambda_{G}$ as before, and  $H_{0}=0$. The mass of the $H$ field reads $M_{H}^{2}=\alpha G_{0}^{2}%
=\alpha\Lambda_{G}^{2}.$ No term $G^{2}H$ is generated,
implying that there is no $s$-channel $H$ production.

Some additional comments are in order:
\begin{enumerate}
    \item The same line of argument can be carried out for $N$ heavy scalar
glueballs $H_{1},\dots, H_{N}$. Only the field $G$ condenses and no $H_{k}G^{2}$ term appears (see also Ref.~\cite{Mueck:2004qg} for a
multiple glueball model). 
\item Here, we have assumed that there is a single dilaton field. We leave for the future the study of the case in which there are two (or more) dilaton fields. Note, however, that such a scenario, even if cannot be excluded a priori, seems quite exotic, since it would imply the emergence of two (or more) distinct energy scales $\Lambda_{G},\Lambda_{H},\dots$ in the low-energy QCD domain.
\item The fact that the $H$ field does not affect $GG$ scattering is true at tree level only. The interaction term proportional to $G^{2}H^{2}$ affects the scattering via the an intermediate $HH$ loop. Yet, since the the mass of an excited scalar glueball is at least $3\gev$~\cite{Morningstar:1999rf}, the loop starts
to be relevant at $\left(  2m_{H}\right)^{2}\sim36 \gevsq$, surely negligible close to threshold.f
\item Relaxing the $Z_2$ symmetry yields a more complicated potential, where multiple scales are generated. A dedicated study of this system is also left for the future. 
\end{enumerate}

\subsection{Non-scalar glueballs}

Here we investigate if glueballs with different quantum numbers can also affect the
results. This question can be answered by a systematic study of the allowed
cubic vertices of the type $G^{2}X$, where $X$
refers to an heavy gluonium. Glueballs with spin $J \ge 2$ are forbidden by dilatation invariance, so we restrict our analysis to all possible spin $0$ and $1$. The list is summarized in Table~\ref{tab:sistem}, and show that it is impossible to accommodate most of the spin parities. 

For example, let us consider the pseudoscalar glueball ($\Tilde{G}$, $J^{PC}=0^{-+}$), see
e.g. Refs.~\cite{Rosenzweig:1981cu,Rosenzweig:1982cb,Eshraim:2012jv,Masoni:2006rz}. The corresponding dilatation and
parity invariant potential reads%

\begin{equation}
V(G,\Tilde{G})=\xi G^{2}\Tilde{G}^{2}+\rho\Tilde{G}^{4}\text{.}%
\end{equation}
Since $\tilde{G}$ cannot condense because of parity conservation, no
$G^{2}\tilde{G}$ or $G^{3}\tilde{G}$---which reduces to the former upon condensation of $G$--- is generated. Similarly, no additonal $G^2$ term is produced, hence the mass of the scalar glueball is unchanged. Similarly, one can discuss the other cases, that break parity or charge conjugation.
The only nontrivial case is given by the oddball $X_\mu$ with $J^{PC}=1^{-+}$, for which the two symmetries are satisfied. 
However, two scalar glueball cannot couple to a spin one state because of Bose symmetry: the vector current for a neutral scalar
\begin{equation}
    G\partial^\mu G - (\partial^\mu G)G = 0
\end{equation}
vanishes identically indeed. 
We conclude that nonscalar glueballs can be  safely
neglected when evaluating $GG$ scattering close to threshold.
\begin{table}[t]
    \centering
    \begin{tabular}{c|c c} \hline
         &  $P$  &  $C$  \\ \hline
    
      $0^{-+} $    & \xmark & \cmark \\

      $0^{+-} $    & \xmark & \xmark \\ 
     
      $0^{--} $    & \cmark & \xmark \\ 
     
      $1^{++} $    & \xmark & \cmark \\ 
     
      $1^{-+} $   &   \cmark & \cmark \\ 
     
      $1^{+-} $    & \xmark & \xmark \\ 
     
     $1^{--} $   & \checkmark & \xmark \\ \hline
    \end{tabular}
    \caption{Symmetries for $XG^2$ terms involving a heavy nonscalar glueball $X$ and two scalar glueballs $G$.  }
    \label{tab:sistem}
\end{table}

\section{Conclusions}
\label{sec:conclusions}
We have investigated the scattering of two  scalar
glueballs in the context of the dilaton potential. 
Using a unitarization prescription, we calculated the
scattering amplitudes for the three lowest nonvanishing waves. These can be computed rigorously in pure gauge Lattice simulations, allowing for a numerical verification of the
validity of the dilaton potential and for an independent determination of its
two parameters, the glueball mass and ---most notably--- the scale
$\Lambda_{G}$ that parametrizes the breaking of the trace anomaly at the
composite mesonic level. The latter is relevant not only for the Yang-Mills
sector, but for numerous low-energy effective models of QCD.\\
Another outcome is the emergence of a
bound state of two scalar glueballs, that we call glueballonium. We have
found that the formation of the glueballonium is possible if the attraction is
strong enough, i.e. if the ratio $\Lambda_{G}/m_{G}$ is smaller than a certain
critical value. For $m_{G}\simeq 1.7\gev$, this is $\Lambda_{G,\text{crit}}\simeq 0.504\gev$. For $\Lambda
_{G}\simeq0.4\gev$ (obtained by matching with the lattice measurement of the gluon condensate), a glueballonium with a mass of about $3.4\gev$ forms. 

If the glueballonium investigated here was confirmed on the lattice, it would also represent an interesting challenge for experimental searches, if not too broad. 
A necessary condition for it to be found experimentally
is that the lightest glueball itself should not be too broad. The most relevant decay
channel of the glueballonium is most likely the one into four  pseudoscalars, that
take place when each constituent glueball undergoes a two-body decay.
Moreover, the simpler decay into two pseudoscalars, that
emerges when the two constituent glueballs scatter into two pseudoscalars
(for instance via one-pion exchange), are also expected to be nonnegligible.
The evaluation of these decays in an important task for the future, if the
existence of this object shall be confirmed by other models and lattice YM.
Moreover, the mass of the glueballonium would be in the energy range covered by the
planned Panda experiment~\cite{Lutz:2009ff,Belias:2020zwx}.
Glueballs can be
also investigated (directly or indirectly via decays of quarkonia) in a
variety of ongoing experiments~\cite{Hamdi:2019dbr,Gutsche:2016wix,Ryabchikov:2019rgx,Marcello:2016gcn,Bediaga:2018lhg,Belle-II:2018jsg}. 

As an additional important outlook, we mention the use of different unitarization schemes, as
well as studying the next-to-leading order of the dilaton potential. 

\begin{acknowledgments}
We thank  Phillip Lakaschus and Alexander K. Nikolla for their contribution in the early stages of this work. We also thank Marc Wagner for useful discussions. F.G. acknowledges financial support from the Polish National Science Centre NCN through the OPUS projects no. 2019/33/B/ST2/00613. 
E.T. acknowledges financial support through the project AKCELERATOR ROZWOJU
Uniwersytetu Jana Kochanowskiego w Kielcach (Development Accelerator of the
Jan Kochanowski University of Kielce), co-financed by the European Union
under the European Social Fund, with no. POWR.03.05.00-00-Z212/18.
\end{acknowledgments}

\appendix
\section{Decays of the scalar glueball}
\label{app:decays}
We consider a toy model where, besides the dilaton/glueball field,  a
scalar meson field $\sigma$ and a triplet pion $\bm{\pi}$
are taken into account. The chiral and dilation invariant potential reads%
\begin{equation}
V(G,\sigma,\bm{\pi})=V(G)+aG^{2}(\sigma^{2}+\bm{\pi}^{2}%
)+\frac{\lambda}{4}(\sigma^{2}+\bm{\pi}^{2})^{2}.\label{toy}%
\end{equation}
Chiral transformations are $O(4)$ rotations in the $(\sigma,\bm{\pi
})$ space. Note that the parameters $a$ and $\lambda$ are dimensionless, thus the
only dimensionful parameter of the model is $\Lambda_{G}$ in the dilaton
potential. Spontaneous symmetry breaking  is realized for $a<0$, when both
fields $G$ and $\sigma$ condense. In principle, one should search for the
minimum in the $\sigma$-$G\ $space, yet for illustrative purposes we neglect
the $G$-$\sigma$ mixing and set the v.e.v. of $G$ as $G_{0}=\Lambda_{G}.$

Then, for $a<0$ the field $\sigma$ develops a v.e.v. for
\begin{equation}
\sigma_{0}^{2}=-2\frac{a}{\lambda}\Lambda_{G}^{2}\simeq f_{\pi}^{2}\text{ ,}%
\end{equation}
where $f_{\pi}$ is the pion decay constant. Moreover, the mass of the scalar
$\sigma$ particle is
\begin{equation}
m_{\sigma}^{2}=2\lambda\sigma_{0}^{2}.
\end{equation}
The coupling of $G$ to pions is given by the term proportional to
$G\bm{\pi}^{2}$ that reads
\begin{equation}
-2a\Lambda_{G}G\bm{\pi}^{2}=\frac{m_{\sigma}^{2}}{2\Lambda_{G}%
}G\bm{\pi}^{2}\,,
\end{equation}
hence the decay $G\rightarrow\pi\pi$ takes the form (inserting a nonzero pion mass):%
\begin{equation}
\Gamma_{G\rightarrow\pi\pi}=6\frac{\sqrt{\frac{m_{G}^{2}}{4}-m_{\pi}^{2}}%
}{8\pi m_{G}^{2}}\left(  \frac{m_{\sigma}^{2}}{2\Lambda_{G}}\right)  ^{2}.
\end{equation}
For $m_{G}\simeq1.7\gev$, $m_{\sigma}\simeq1.3\gev$ (roughly corresponding to the
$f_{0}(1370)$), and $\Lambda_{G}\simeq0.4\gev$,  one gets $\Gamma_{G\rightarrow\pi\pi
}\simeq0.310\gev$. If one extends to $SU(3)$,  get
\begin{align}
\Gamma_{G\rightarrow KK}&=8\frac{\sqrt{\frac{m_{G}^{2}}{4}-m_{K}^{2}}}{8\pi
m_{G}^{2}}\left(  \frac{m_{\sigma}^{2}}{2\Lambda_{G}}\right)  ^{2}\,, & \Gamma_{G\rightarrow\eta\eta}&=2\frac{\sqrt{\frac{m_{G}^{2}}{4}-m_{\eta}^{2}}%
}{8\pi m_{G}^{2}}\left(  \frac{m_{\sigma}^{2}}{2\Lambda_{G}}\right)  ^{2}\,.
\end{align}
Numerically, $\Gamma_{G\rightarrow KK}\simeq0.340\gev$ and $\Gamma_{G\rightarrow\eta\eta}\simeq0.080\gev$. The sum of the 3
pseudoscalar channels amounts $0.729\gev$. Furthermore, one needs to consider at least  $G\rightarrow\rho\rho\rightarrow4\pi$, which is expected to be
sizable, but cannot be determined within this simple approach. Such a glueball
would have a total decay of about $1\gev$ and would not be observable. We recall also that
$\Lambda_{G}\simeq0.4\gev$ is on the right side of the interval: the smaller $\Lambda_{G}$, the larger the decay widths. Conversely, widths get
smaller if $m_{\sigma}$ is taken to be smaller.

Finally, we check the dependence of the parameters with the
number of colors $N_{c}$. Besides the scaling $\Lambda_{G}\propto N_{c}$ and
$m_{G}\propto N_{c}^{0}$ already encountered in Sec.~\ref{sec:theory}, we have
$\lambda\propto N_{c}^{-1}$ (the $\pi\pi$ scattering amplitude scales as
$N_{c}^{-1}$ as expected for meson scattering), while
$a\propto N_{c}^{-2}$, since it describes $GG\to \pi\pi$ scattering~\cite{Witten:1979kh,Lebed:1998st}. It then follows that $\sigma_{0}$ scales as
$N_{c}^{1/2}$~\cite{Heinz:2011xq} and $m_{\sigma}$ is $N_{c}$-independent, as expected for the masses of mesons and glueballs. The decay width $\sigma\rightarrow\pi\pi$ scales as
$N_{c}^{-1}$, since the amplitude is proportional to $\lambda\sigma_{0}\propto
N_{c}^{-1/2}.$ Finally, each decay of the glueball into two pseudoscalar
mesons behaves as $N_{c}^{-2}$, a result which is also in agreement with the literature~\cite{Witten:1979kh,Lebed:1998st}.

\section{Kinematics and conventions}
\label{app:kinematics}
We consider the elastic scattering of two identical scalar glueballs $G(p_1) \, G(p_2) \to G(p_3) \, G(p_4)$.
The conventional Mandelstam variables and the relation to the scattering angle $\theta$ are given by:
\begin{align}
s &= (p_{1} + p_{2})^{2}\,,\\
t  &  =(p_{1}-p_{3})^{2}=-2k^2(1-\cos\theta)\leq0\,,\\
u  &  =(p_{2}-p_{3})^{2}=-2k^2(1+\cos\theta)\leq0\,,
\end{align}
where $k = \frac{1}{2}\sqrt{s - 4m_G^2}$ is the 3-momentum of any particle in the center of mass.
It is immediate to verify that $s+t+u=4m_{G}^{2}$. 

The amplitude $A(s,t,u)$ can be re-expressed as $A(s,\cos\theta)$, and partial waves can be defined by:
\begin{align}
A(s,t,u)&=A(s,\cos \theta)=\sum_{\ell=0}^{\infty}(2\ell+1)A_{\ell}(s)P_{\ell}(\cos\theta)\,,\\
A_{\ell}(s)&=\frac{1}{2}\int_{-1}^{1}d\cos\theta A(s,\cos\theta)P_{\ell}(\cos\theta)\,.\label{l-ampl}%
\end{align}
where $P_{\ell}(\cos\theta)$ are the Legendre polynomials. Bose symmetry imposes $A(s,\cos\theta)$ to be symmetric in $\cos\theta$, so that odd waves vanish. The reduced (kinematical singularity and zero free) amplitudes are given by:
\begin{equation}
    \hat A_{\ell}(s) = \frac{1}{k^{2l}}A_{\ell}(s)\,.
\end{equation}
The amplitudes at threshold are often described in term of scattering lengths, which we normalize to:
\begin{equation}
     a_\ell = \frac{\hat A_\ell(4m_G^2)}{32\pi \,m_G}\,.
\end{equation}

Partial waves can also be parametrized in terms of scattering shift and inelasticity,
\begin{equation}
\frac{\eta_\ell(s) e^{2i\delta_{\ell}(s)}-1}{2i} = \frac{1}{2} \cdot \frac{k^{2\ell +1}}{8\pi \sqrt{s}} \hat A_{\ell}(s) \equiv \rho_\ell(s) \hat A_{\ell}(s), \label{phaseshift_tl}
\end{equation}
hence
\begin{align}
\delta_{\ell}(s) &=\frac{1}{2}\arg\left[  1+2i \rho_\ell(s) \hat A_{\ell}(s)\right]  , \label{eq:phasearg}\\
\eta_\ell(s) &=\left|  1+2i \rho_\ell(s) \hat A_{\ell}(s)\right| . \label{eq:inel}%
\end{align}

If the amplitude is unitarity, then $\eta_\ell(s) = 1$ and 
\begin{equation}
    \im \hat A_{\ell}(s) = \rho_\ell(s) \left|\hat A_{\ell}(s)\right|^2\,.
\end{equation}
We finally recall that the differential and the total cross sections
are:
\begin{align}
\frac{d\sigma}{d\Omega}&=\frac{\left\vert A(s,\cos\theta)\right\vert ^{2}}%
{64\pi^{2}s}\,,\\
\sigma &=\frac{1}%
{64\pi s}\int_{-1}^1 d\cos\theta\,\left|A(s,\cos\theta)\right|^2 = \frac{1}%
{32\pi s}\sum_{\ell =0}^\infty (2\ell+1) k^{4\ell} \left|\hat A_\ell(s)\right|^2.
\end{align}
\section{Generic tree-level scattering length}
\label{app:length}
For $\ell \ge 2$, the partial waves reads
\begin{equation}
    A_\ell(s) = \frac{50m_G^4}{\Lambda_G^2} \frac{Q_\ell\left(1 + \frac{m_G^2}{2k^2}\right)}{2k^2}.
\end{equation}
For $k\to 0$, the argument of the Legendre function diverges. One can use the definition:
\begin{equation}
   \frac{Q_\ell(x)}{2k^2} = \frac{1}{4k^2}\int_{-1}^{1} \frac{P_\ell(z)}{x - z} dz = \frac{1}{4k^2 x}\int_{-1}^{1} P_\ell(z) \sum_{n=0}^\infty \left(\frac{z}{x}\right)^n dz \,.
\end{equation}
The term $z^n$ can be thought of a superposition of Legendre polynomials of degree $\le n$. Thus the terms with $n < l$ will vanish because of orthogonality:
\begin{equation}
   = \frac{1}{4k^2 x}\int_{-1}^{1} P_\ell(z) \sum_{n = \ell}^\infty \left(\frac{z}{x}\right)^n dz \xrightarrow{k \to 0}\left(\frac{2k^2}{m_G^2}\right)^\ell\frac{1}{2m_G^2}\int_{-1}^{1} P_\ell(z) z^\ell dz\,.
\end{equation}
Still because of orthogonality, we can add lower order polynomial to $z^\ell$ without changing the integral, to make it proportional to $P_\ell(z)$, the proportionality given by the coefficient of the leading power of $z$ in the Legendre polynomial:
\begin{equation}
   =\left(\frac{2k^2}{m_G^2}\right)^\ell\frac{2^\ell (\ell!)^2}{2m_G^2 (2\ell)!}\int_{-1}^{1} P_\ell(z) P_\ell(z) dz = \frac{(\ell!)^2}{(2 \ell + 1)(2\ell)!m_G^2 }\left(\frac{4k^2}{m_G^2}\right)^\ell\,.
\end{equation}
The scattering lengths read:
\begin{equation}
a_\ell = \frac{25(\ell!)^2 4^{\ell-2}}{\pi\,(2 \ell + 1)(2\ell)! }\frac{1}{\Lambda_G^2m_G^{2\ell-1}}.\
\end{equation}

\section{Asymptotic behaviour of the phase shifts}
\label{app:asymptotics}

\begin{figure}[t]
\includegraphics[width=.4\textwidth]{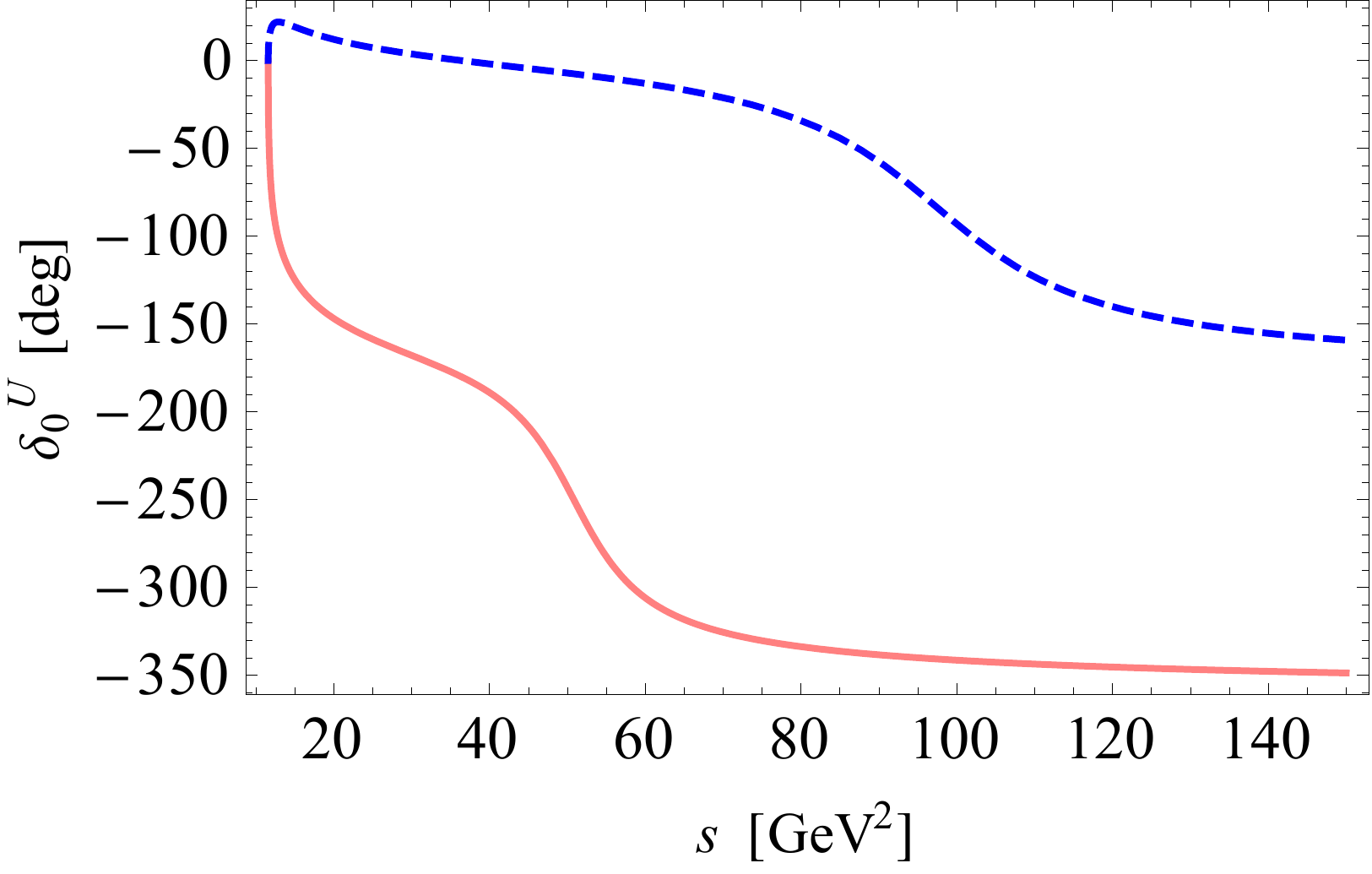} \hspace{.5cm}
\includegraphics[width=.4\textwidth]{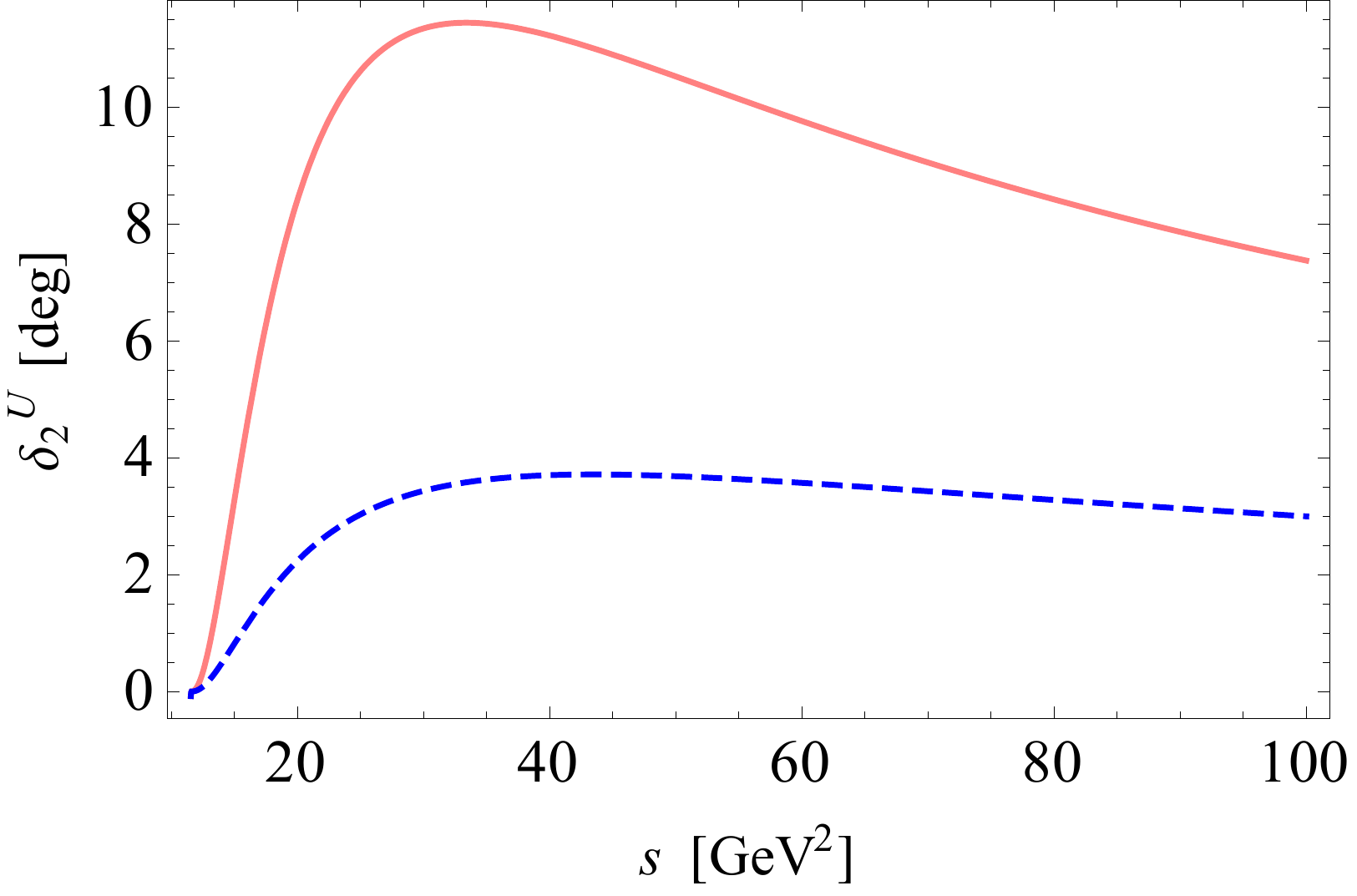}
\caption{Asymptotic behaviour of the unitarized phase shifts as function of $s$ for $\Lambda_G=0.4\gev$ (pink, solid) and $\Lambda_G=0.8\gev$ (blue, dashed). In the two panels we show the $S$- and the $D$-waves.}				\label{fig:8}
\end{figure}

In Fig.~\ref{fig:8} the $S$-wave and the $D$-wave unitarized phase-shifts are shown for different values of $\Lambda_{G}$ and up to large values of $s$. Even though the range plotted is above the inelastic threshold and should not be regarded as physical, these plots show that
the expectations from Levinson's theorem are fulfilled, In particular, when the glueballonium is present, the phase shifts tends to $-360^{\circ},$ otherwise it saturates at $-180^{\circ}$.



\section{Alternative unitarization}
\label{app:noverd}
In this Appendix we investigate another possible unitarization method, in order to estimate the dependence of our results on the adopted scheme. 
To this end, we
consider the well-known $N/D$ unitarization~\cite{Hayashi:1967bjx,Frazer:1969euo} in its simplest
form (e.g. \cite{Cahn:1983vi,Gulmez:2016scm}). At lowest order, the left-hand cut is given by the tree-level amplitude, and the analytic properties are preserved. The $S$-wave
amplitude reads:%

\begin{equation}
A_{0}^{\text{N/D}}(s)=\frac{A_{0}(s)}{D(s)},
\label{AN/D}
\end{equation}
where%
\begin{equation}
D(s)=1-\dfrac{(s-m_{G}^{2})}{\pi}\int_{4m_G^{2}}^{\infty}\frac{\frac{1}{2}%
\frac{\sqrt{\frac{s^{\prime}}{4}-m_{G}^{2}}}{8\pi\sqrt{s^{\prime}}}%
A_{0}(s^{\prime})}{(s^{\prime}-s-i\varepsilon)(s^{\prime}-m_{G}^{2}%
)}ds^{\prime}\label{dofs}%
\text{ .}
\end{equation}
We require that  the unitarized amplitude coincides with the tree-level one in the vicinity of the single-particle pole $s=m_{G}%
^{2}$. In other
words, we \textit{assume} that the tree-level amplitude contains the correct
position as well as the residue of the one-particle pole. 
Admittedly, at this stage this is  an additional convenient assumption driven by the fact that, besides the single particle pole, there are no other constraints from lattice data that could be imposed.
Moreover, one
subtraction is sufficient, since no ghost appears. The extension to
higher waves is straightforward, but we focus on the $S$-wave below threshold in order to study the emergence of the glueballonium. 
Note, following  Ref.~\cite{Cahn:1983vi}, the single-particle pole is left in the numerator.

\begin{figure}[t]
\includegraphics[width=.45\textwidth]{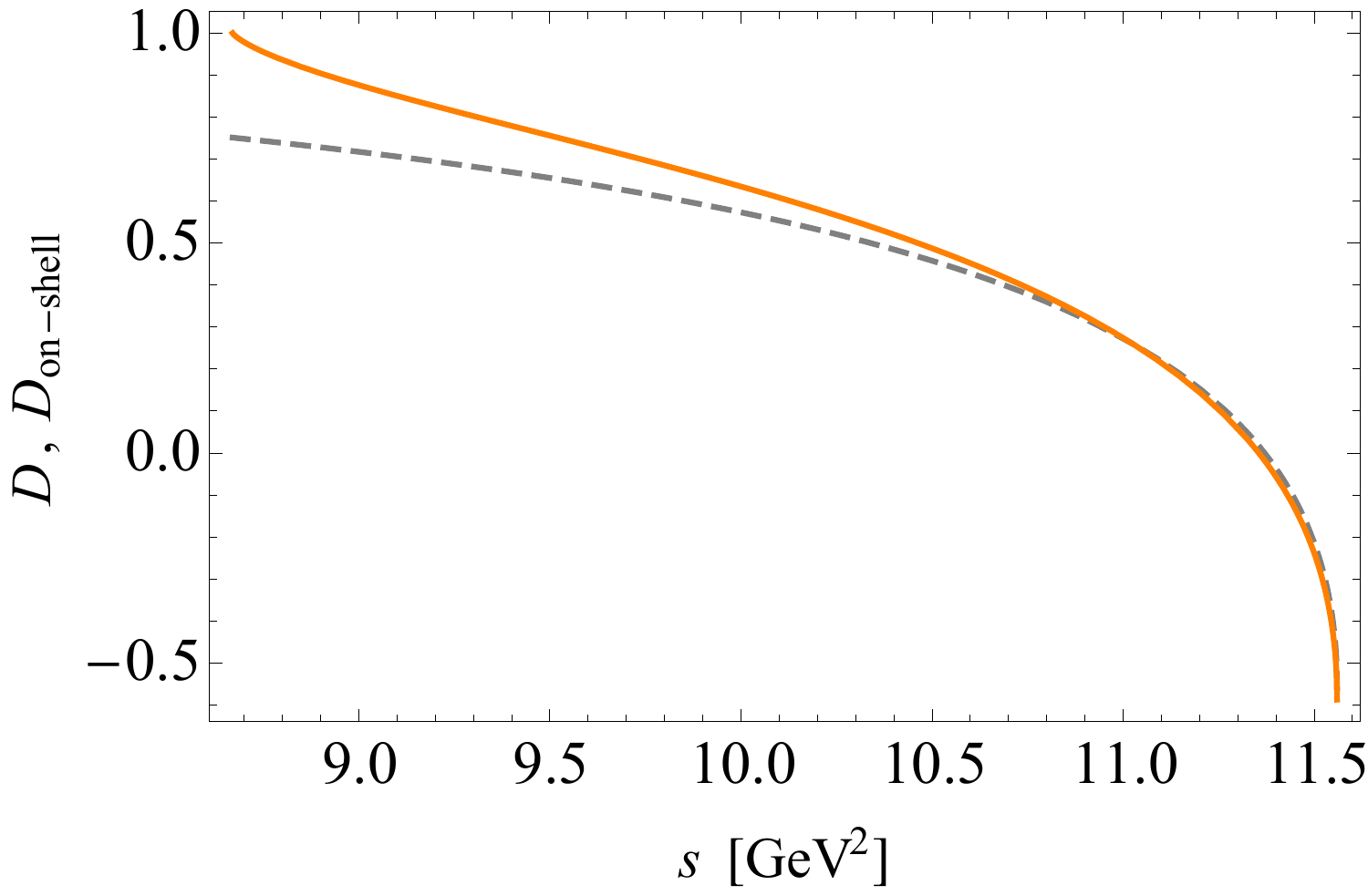} \hspace{.5cm}
\includegraphics[width=.45\textwidth]{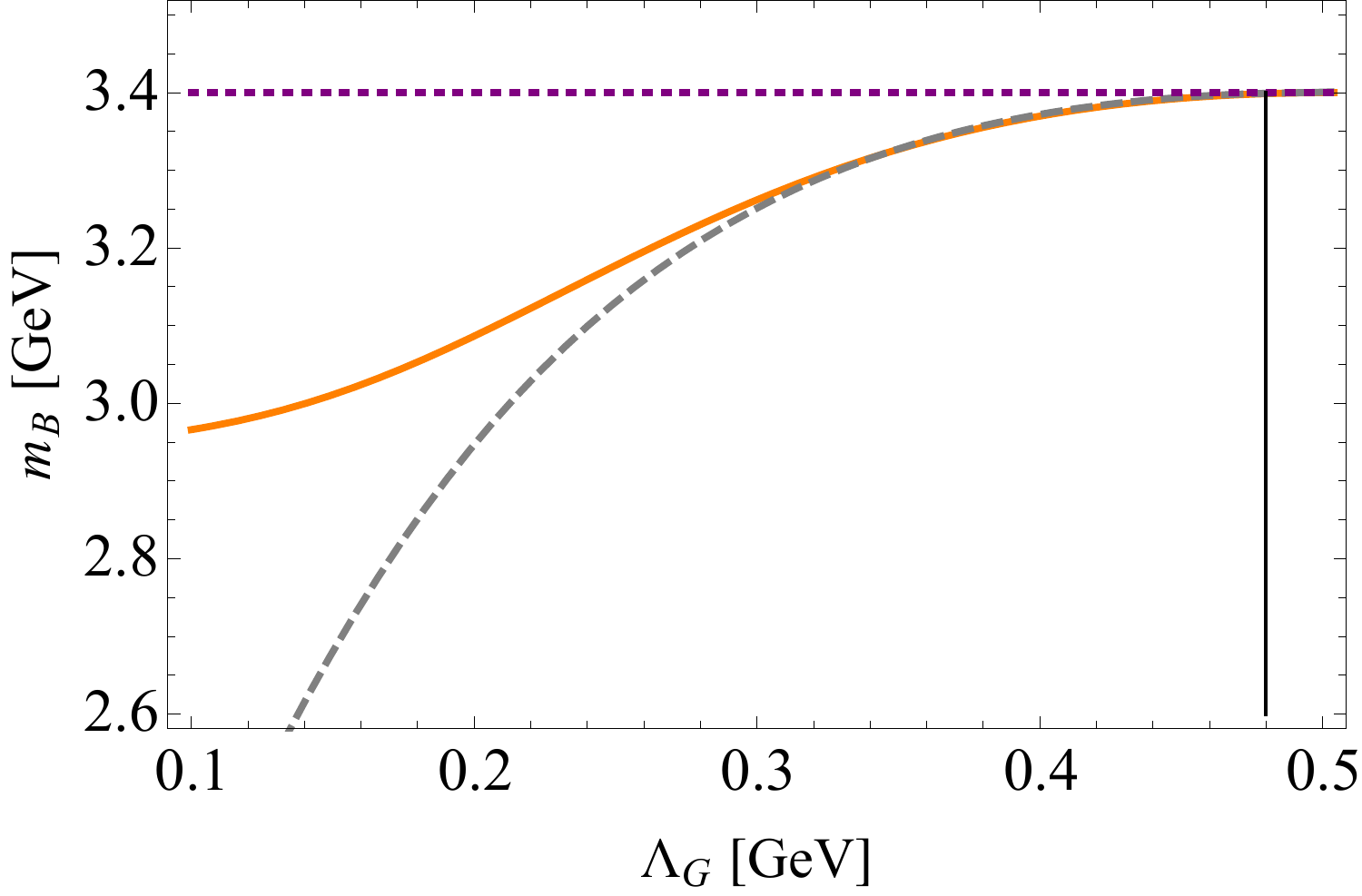}
\caption{Left panel: Comparison of the denominators of the unitarized amplitudes: $D(s)$  (gray,  dashed) of Eq.~\eqref{dofs} for the $N/D$ case and $D_{\text{on-shell}}(s)$ (orange, continuous) of Eq.~\eqref{Donshell} for the on-shell scheme. Right panel: The mass of the glueballonium as function of the parameter $\Lambda_{G}$ with the on-shell  (orange, continuous) and $N/D$ (gray,  dashed) unitarizations. 
}				\label{fig:9}
\end{figure}

The bound-state equation is given by $D(s)=0$ for $s < 4m_G^2$. Although this equation allows in principle to study bound states even below the left-hand branch point $s < 3 m_G^2$, results are trustworthy only above it.
The corresponding critical value of $\Lambda_{G}$ is $\Lambda_{G,crit}=0.500\gev$, which is very similar to the value $\Lambda_{G,\text{crit}}=0.504\gev$ obtained by
the unitarization of Sec.~\ref{sec:unitarization}. For the illustrative value $\Lambda_{G}=0.4\gev$
the glueballonium mass reads $3.372\gev$, in good agreement
with the value reported in the main text. For comparison, we plot in Fig.~\ref{fig:9} 
(left panel) the denominator $D(s)$ of Eq.~\eqref{dofs} as well as the analogous quantity
of the on-shell unitarization of Eq.~\eqref{eqaluni}:
\begin{equation}
 D_{\text{on-shell}} =   1-A_{0}(s)\Sigma(s)
 \text{ .}
 \label{Donshell}
\end{equation}
The functions display a similar qualitative behavior and are also numerically very close in the vicinity of the threshold. 

In Fig.~\ref{fig:9} (right panel) 
we show the behavior of the mass of the glueballonium for the two
unitarization schemes. They are indeed very similar for $\Lambda_{G}%
\gtrsim0.3\gev$ and they depart from each other for lower values. Yet, since
the phenomenological expected range for the parameter $\Lambda_{G}$ is larger
than $0.3\gev$, we can be cautiously confident that our results are not
strongly affected by the details of the employed unitarization.
As stated above, for small $\Lambda_{G}$ (strong attraction) the unitarization approaches deliver different results. In particular, while in the on-shell unitarization of Sec.~\ref{sec:unitarization}
the mass of the bound state cannot be smaller then
$\sqrt{3}m_{G}$ (reached formally for $\Lambda_{G}\rightarrow0$), in the $N/D$ scheme the limit is $m_{G}.$ 
Of course, this difference has no physical relevance, as these approximated methods are not realistic when the attraction is too large.

In conclusion, even though also the present $N/D$ approach is subject to certain ad-hoc assumptions, the fact that similar results for the emergence of a bound state are obtained by this method and by the twice-subtracted on-shell approach discussed in the main text, can be regarded as a hint about the consistency of our results.

\bibliographystyle{apsrev4-2.bst}
\bibliography{quattro}
\end{document}